\documentclass[lettersize,journal]{IEEEtran}
\usepackage{amsmath,amsfonts,amssymb}
\usepackage{algorithmic}
\usepackage{algorithm}
\usepackage{array}
\usepackage{textcomp}
\usepackage{stfloats}
\usepackage{lipsum}  
\usepackage{url}
\usepackage{verbatim}
\usepackage{comment}
\usepackage{subcaption}
\usepackage{color}
\usepackage{graphicx}
\usepackage{cite}

\newcommand{\p}{\mathrm{p}}
\newcommand{\PP}{\mathrm{P}}
\newcommand{\E}{\mathrm{E}}
\newcommand{\e}{\mathrm{e}}
\newcommand{\Var}{\mathrm{Var}}

\newcommand{\Cov}{\mathrm{Cov}}

\newcommand{\bin}{\mathrm{Bin}}
\makeatletter
\newcommand{\vastt}{\bBigg@{3}}
\newcommand{\vast}{\bBigg@{4}}
\newcommand{\Vast}{\bBigg@{5}}

\makeatother

\begin{document}

\title{Ratio Shift Keying Modulation for Time-Varying Molecular Communication Channels}
\author{
        M. Okan Araz*,
        Ahmet R. Emirdagi*, M. Serkan Kopuzlu*,
        and Murat Kuscu
        \thanks{*These authors contributed equally.}
       \thanks{The authors are with the Nano/Bio/Physical Information and Communications Laboratory (CALICO Lab), Department of Electrical and Electronics Engineering, Koç University, Istanbul, Turkey (e-mail: \{maraz18, aemirdagi18, mkopuzlu18,  mkuscu\}@ku.edu.tr).}
	   \thanks{This work was supported in part by the European
Union’s Horizon 2020 Research and Innovation Programme through the Marie Skłodowska-Curie Individual Fellowship under Grant Agreement 101028935, and by The Scientific and Technological Research Council of Turkey (TUBITAK) under Grant \#120E301.}
\thanks{An earlier version of this work has been presented at ACM NanoCom'22, Barcelona, Spain \cite{10.1145/3558583.3558845}. }
}

\maketitle

\begin{abstract}
Molecular Communications (MC) is a bio-inspired communication technique that uses molecules to encode and transfer information. Many efforts have been devoted to developing novel modulation techniques for MC based on various distinguishable characteristics of molecules, such as their concentrations or types. In this paper, we investigate a particular modulation scheme called Ratio Shift Keying (RSK), where the information is encoded in the concentration ratio of two different types of molecules. RSK modulation is hypothesized to enable accurate information transfer in dynamic MC scenarios where the time-varying channel characteristics affect both types of molecules equally. To validate this hypothesis, we first conduct an information-theoretical analysis of RSK modulation and derive the capacity of the end-to-end MC channel where the receiver estimates concentration ratio based on ligand-receptor binding statistics in an optimal or suboptimal manner. We then analyze the error performance of RSK modulation in a practical time-varying MC scenario, that is mobile MC, in which both the transmitter and the receiver undergo diffusion-based propagation. Our numerical and analytical results, obtained for varying levels of similarity between the ligand types used for ratio-encoding, and varying number of receptors, show that RSK can significantly outperform the most commonly considered MC modulation technique, concentration shift keying (CSK), in dynamic MC scenarios.
\end{abstract}


\section{Introduction}
\IEEEPARstart{B}{io-inspired} Molecular Communications (MC), which relies on biochemical molecules to encode and exchange information, is promising for interconnecting heterogeneous bio-nano things, e.g., engineered bacteria and nanobiosensors, thereby enabling unprecedented healthcare applications, such as intrabody continuous health monitoring within the Internet of Bio-Nano Things (IoBNT) framework \cite{akyildiz2015internet,kuscu2021internet, akan2016fundamentals}. Over the last 15 years, there has been significant research interest in theoretical aspects of MC, such as channel modeling, detection and modulation techniques \cite{kuscu2019transmitter}. More recently, experimental studies have started to accompany this theoretical body of work \cite{kuscu2021fabrication, grebenstein2019molecular}. 

As the nature of information carriers in MC, i.e., molecules, is fundamentally different than that of the electromagnetic (EM) waves utilized in conventional communication technologies, researchers have developed novel modulation techniques that can exploit the distinguishable properties of molecules, such as concentration (concentration-shift-keying - CSK) \cite{kuran2020survey}, molecule type (molecule-shift-keying - MoSK) \cite{kuran2012interference}, and release time of molecules from the transmitter (release-time-shift-keying - RTSK) \cite{murin2016communication}. Relatively less interest has been devoted to exploiting the concentration ratio between different types of molecules released simultaneously. This so-called ratio-shift-keying (RSK) modulation was first investigated in \cite{kim2013novel} as part of a large set of MC modulation techniques exploiting the properties of isomers. Accordingly, the authors proposed encoding information into the concentration ratio of transmitted isomers that differ in the arrangement of constituent monomers. However, except for this initial investigation of the isomer-based RSK modulation scheme, a thorough numerical performance analysis of RSK for practical MC scenarios is absent in the current literature.

\begin{figure}
	\includegraphics[width=\linewidth]{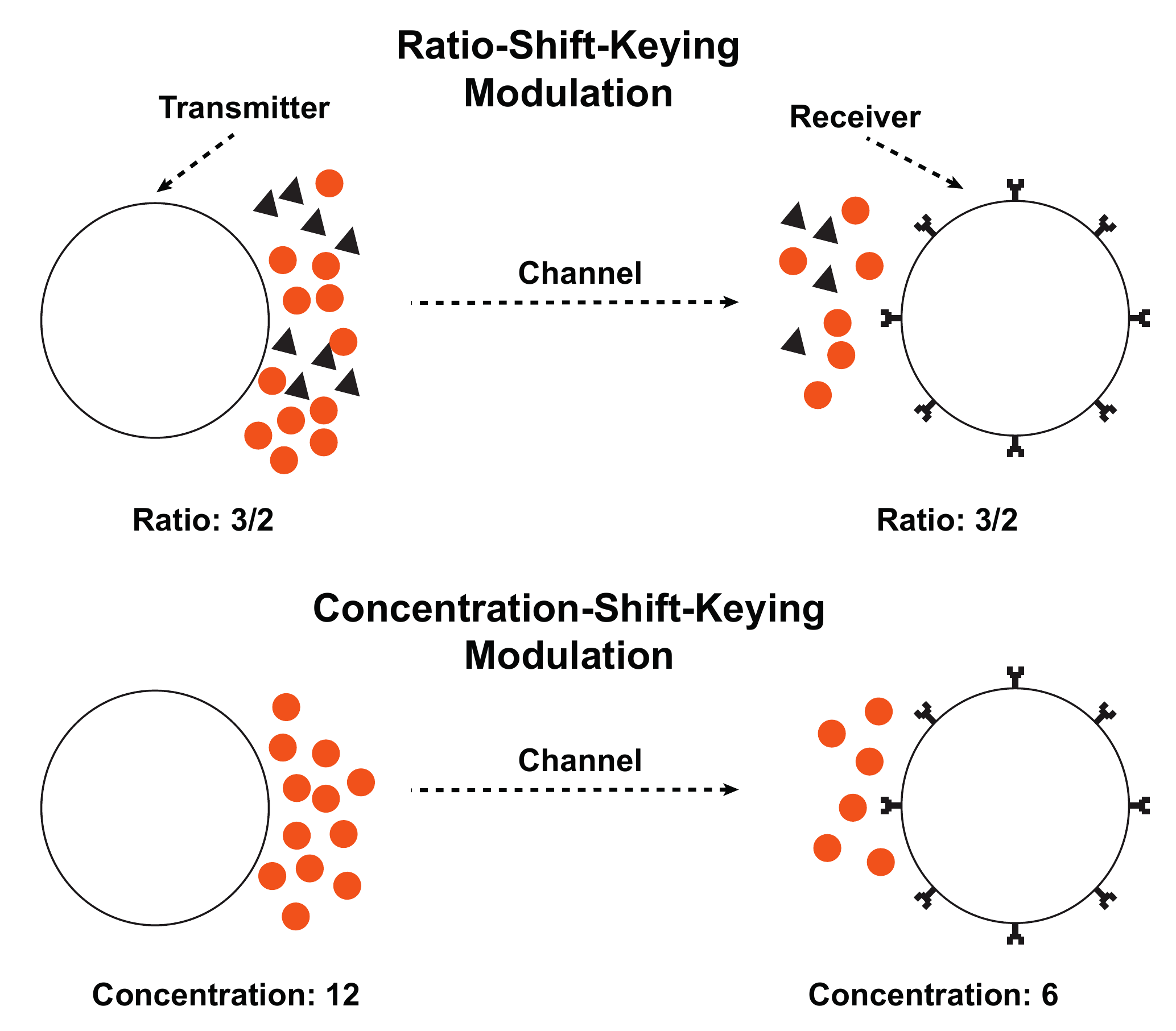}
	\caption{Demonstration of ligand transmission from transmitter to receiver in a time-varying MC channel using RSK and CSK. Even though the concentration of ligands changes, the concentration ratio between these two ligands is preserved since both types of ligands are equally affected by the variations in CIR.}
	\label{fig:rsk_channel}
\end{figure}

RSK can have significant advantages over other MC modulation techniques under certain conditions of the MC channel and the MC transceivers. First, the same concentration ratios encoding a particular symbol set can be obtained with different absolute concentrations of individual types of molecules, offering extended opportunities for energy-efficient (i.e., molecule-efficient) information exchange. Second, RSK can be more robust against dynamic variations in transmit power and channel impulse response (CIR) if the effect of these variations is molecule-type invariant. This can be exemplified by the mobile MC case where the diffusion coefficients of different types of molecules are equal. In that case, the time-varying CIR due to the mobility of the transceivers would be the same for both types of molecules at all times, preserving the concentration ratio in the received signal as demonstrated in Fig. \ref{fig:rsk_channel}. Similarly, RSK can be relatively robust in cases where the channel has enzymes that degrade both types of molecules at the same rate, which would not alter the received concentration ratio. We can also exemplify the potential advantages of RSK by considering cases where the transmitter, with a finite reservoir of molecules or fluctuating molecule generation or harvesting mechanisms, which can result in time-varying transmission profiles in terms of the absolute number of transmitted molecules. If the transmitter is able to maintain the transmitted concentration ratios under such conditions, RSK can preserve its reliability. 
 
All of the aforementioned advantages of RSK, however, are contingent upon the ability of the MC receiver to accurately detect the transmitted concentration ratios. In this paper, we investigate the performance of RSK modulation for both stationary and mobile transmitter-receiver scenarios, considering a physically-relevant MC receiver architecture that is equipped with \textbf{a single type of ligand receptors} interacting with different types of information molecules (i.e., ligands) in a \textbf{cross-reactive} manner. By exploiting the difference in the affinity of the different types of ligands with the receptors, which is reflected in the difference in receptor-ligand bound time duration statistics, the receiver is able to estimate the received concentration ratio in a maximum-likelihood (ML) manner \cite{kuscu2019channel}. However, due to the complexity of this optimal ML estimation scheme, we also consider a practical and suboptimal estimation method based on the biological kinetic proof-reading (KPR) mechanism \cite{kuscu2021detection}.

To evaluate the performance of RSK modulation, we first conduct an information theoretical analysis and analytically derive the approximate capacity of an end-to-end MC channel and the corresponding optimal input distribution with a receiver performing either optimal or suboptimal ratio estimation. The numerical results obtained by varying system parameters, such as the similarity between ligand types used for modulation and the number of receptors, are compared to the capacity of the MC channel using the more conventional CSK modulation. In the second part of our analysis, we evaluate the error performance of both RSK and CSK modulation in a practical mobile MC scenario, where both the transmitter and receiver are mobile. We analytically derive the symbol error probability (SEP) for both modulation schemes which are then compared to the numerical results obtained via Monte Carlo simulations. In addition to the system configurations examined in the information-theoretical analysis, we also investigate the effects of the diffusion coefficient of the transmitter-receiver pair on the performance of both modulation schemes in our analysis of mobile MC.

The results of the information theoretical analysis in the first part show that an end-to-end MC channel with RSK manifests similar capacity as CSK, but significantly outperforms CSK if the transmit power is limited. The performance of the suboptimal estimator, which is more applicable to biological mechanisms due to its low computational complexity, is revealed to be quite close to that of the optimal estimator. These results indicate the potential of RSK in time-varying channel and transceiver conditions, and hint at its advantages over CSK and potentially other modulation techniques for the design of energy-efficient (i.e., molecule-efficient) MC systems. The results of the error performance analysis in the second part demonstrate the potential of RSK when both receiver and transmitter are mobile in a time-varying end-to-end MC channel. Numerical results show that RSK has a better error performance than CSK in mobile MC cases, which becomes more prominent as the mobility of transmitter and receiver increases.  As a general conclusion, MC channel with RSK has a similar capacity as the one with CSK, however, in time-varying MC scenarios, such as mobile MC, RSK outperforms CSK when the CIRs for both type of ligands are affected equally. 

The remainder of this paper is organized as follows. In Section II, we provide a brief overview of the statistics of ligand-receptor binding reactions. In Section III, we present the mathematical framework for the concentration ratio and concentration estimation from ligand-receptor binding statistics. In Section IV, we introduce the MC model setting used in the derivations and analyses in subsequent sections. We derive and evaluate the information-theoretical MC channel capacity with RSK and CSK modulation in Section V. In Section VI, we present the error performance analysis of RSK and CSK for a practical mobile MC scenario. Lastly, we conclude the paper in Section VII.

\section{Statistics of Ligand-Receptor Binding Reactions}
\label{sec:statistics}
Ligand-receptor interactions are key to communication and sensing in nature, as most biological cells, e.g., most bacteria, T-cells, express surface receptors as selective biorecognition elements, which undergo reversible reactions with specific types of molecules \cite{bialek2012biophysics}. These interactions are then translated into molecular representations inside the cell, which in turn, inform the cell's subsequent actions. On the other hand, MC literature has so far mostly focused on receiver architectures that neglect the presence of receptors and ligand-receptor interactions. However, recent studies highlighted the significant impact of these interactions on the overall MC performance, and hinted at unique opportunities that can be obtained from their statistics \cite{kuscu2021fabrication, kuscu2016physical}. 

Ligand-receptor binding interactions, if monovalent, can be formulated by a two-state continuous-time stochastic process with the states corresponding to the bound (B) and unbound (U) states of the receptor: 
\begin{equation}
\label{equilibrium}
    \text{U} \underset{k^{-}}{\stackrel{c_{L}(t) k^{+}}{\rightleftharpoons}} \text{B}
\end{equation}
where, $c_L(t)$ is the time-varying ligand concentration in the vicinity of the receptor, $k^+$ and $k^-$ are the binding and unbinding rates of the ligand-receptor pair, respectively. 

Due to the low-pass characteristics of the diffusion-based MC channel, the bandwidth of $c_L(t)$ is typically significantly smaller than the characteristic frequency of the binding reactions, i.e., $f_B=c_L(t ) k^+ +k^- $ \cite{kuscu2019channel}. Hence, ligand-receptor reactions can be assumed to be in equilibrium with a stationary ligand concentration in a time window of interest, and $c_L(t)$ can be simplified to $c_L$. Under these equilibrium conditions, the process \eqref{equilibrium} can be represented by a continuous-time Markov process (CTMP), and the probability of a receptor being in the bound state is given as follows:
\begin{equation}
\label{eq:success_prob_csk}
   \p_B=\frac{c_L}{c_L + K_D },
\end{equation} where $K_D =k^-/k^+$ is the dissociation constant, which is inversely proportional to the ligand-receptor binding affinity. Considering that there are $N_R$ number of receptors that do not interact with each other, and are exposed to the same ligand concentration, the number of bound receptors can be expressed as a binomial distribution, $n_B \sim \bin(\p_B,N_R)$. Following from the memoryless property of the CTMP, the bound and unbound times of the receptors at equilibrium are exponentially distributed with the rate parameters depending on the binding and unbinding rates of the ligand-receptor pair, respectively. 

In the case of two different types of ligands in the receptors' vicinity, both ligands can bind to the same receptors, but with different affinities, i.e., different $K_D$, which are reflected to the bound state probability of the receptors as follows
\begin{equation}
   \p_B=\frac{c_1/K_{D,1} + c_2/K_{D,2}}{1 + c_1/K_{D,1} + c_2/K_{D,2} },
   \label{eq:prob_binding_mixture}
\end{equation} where $c_1$ and $c_2$ are the concentrations of type-1 and type-2 ligands whose dissociation constants are denoted by $K_{D,1}$ and $K_{D,2}$ respectively. Due to the interchangeability of the summands, \eqref{eq:prob_binding_mixture} cannot be used to estimate the individual ligand concentrations, $c_1$ and $c_2$. As a result, in cases where different ligand types coexist in the channel, necessary statistics regarding individual ligand concentrations can only be obtained by analyzing the continuous history of ligand binding and unbinding events over receptors.

In diffusion-limited cases, the characteristic rate of diffusion is much smaller than the ligand-receptor binding reaction rates, which allows for the simplification of the binding rates for circular receptors as $k^+ = 4 Da$, with $D$ and $a$ denoting the diffusion constant of molecules and the effective receptor size, respectively \cite{mora2015physical}. Assuming that the size difference between different ligand types is negligible, the diffusion constant, which then only depends on the temperature and viscosity of the channel medium, can be assumed to be equal for all ligand types. Under these assumptions, the probability of observing a particular bound time duration $\p\left(\tau_{b}\right)$ in a single receptor can be written as a mixture of exponential distributions:
\begin{align}\label{eq:prob_taub}  
\p(\tau_{b})   =  \sum_{j=1}^2 \alpha_j k_j^- \e^{-k_j^- \tau_{b}} .
\end{align} where $ \alpha_j = c_j /c_{tot}$ is the concentration ratio of the j$th$ ligand, and $c_{tot}$ is the total ligand concentration. Then, the log-likelihood function for observing a set of bound time durations over $N_R$ independent receptors can be written as
\begin{equation}
    \mathcal{L}(\{\tau_b\}|\alpha) = \sum_{i=1}^{N_R} \ln \p(\tau_{b,i}),
    \label{eq:likelihood_boundtimedurations}
\end{equation} where $\tau_{b,i}$ is the bound time duration observed on the $i^\text{th}$ receptor. 

\section{Parameter Estimation based on Ligand Receptor Binding Statistics}

\subsection{Optimal Estimation of the Ligand Concentration Ratios}
\label{sec:optimal}
The optimal estimation of ligand concentration ratios can be obtained using an ML approach by setting the first derivative of the likelihood function for bound time durations \eqref{eq:likelihood_boundtimedurations} with respect to the concentration ratio of type-1 ligands, i.e., $\alpha$, to zero: 
\begin{equation}
\label{eq:ml_estimation}
    \sum_{j=1}^{N_R} \frac{k^-_1 e^{- k^-_1 \tau_{b,j}}}{ \alpha k^-_1 e^{- k^-_1 \tau_{b,j}} +(1-\alpha) k^-_2 e^{- k^-_2 \tau_{b,j}}} = 0.
\end{equation}
However, solving this equation for the optimal ratio estimation requires the use of computationally complex algorithms, which may not be feasible for resource-constrained bio-nano devices. As an alternative, we will investigate a practical and suboptimal concentration ratio estimation scheme which can be implemented by biological circuits \cite{kuscu2019channel}.  

\subsection{Suboptimal Estimation of the Ligand Concentration Ratios}
\label{sec:suboptimal_est}
\begin{figure}
	\includegraphics[width=\linewidth]{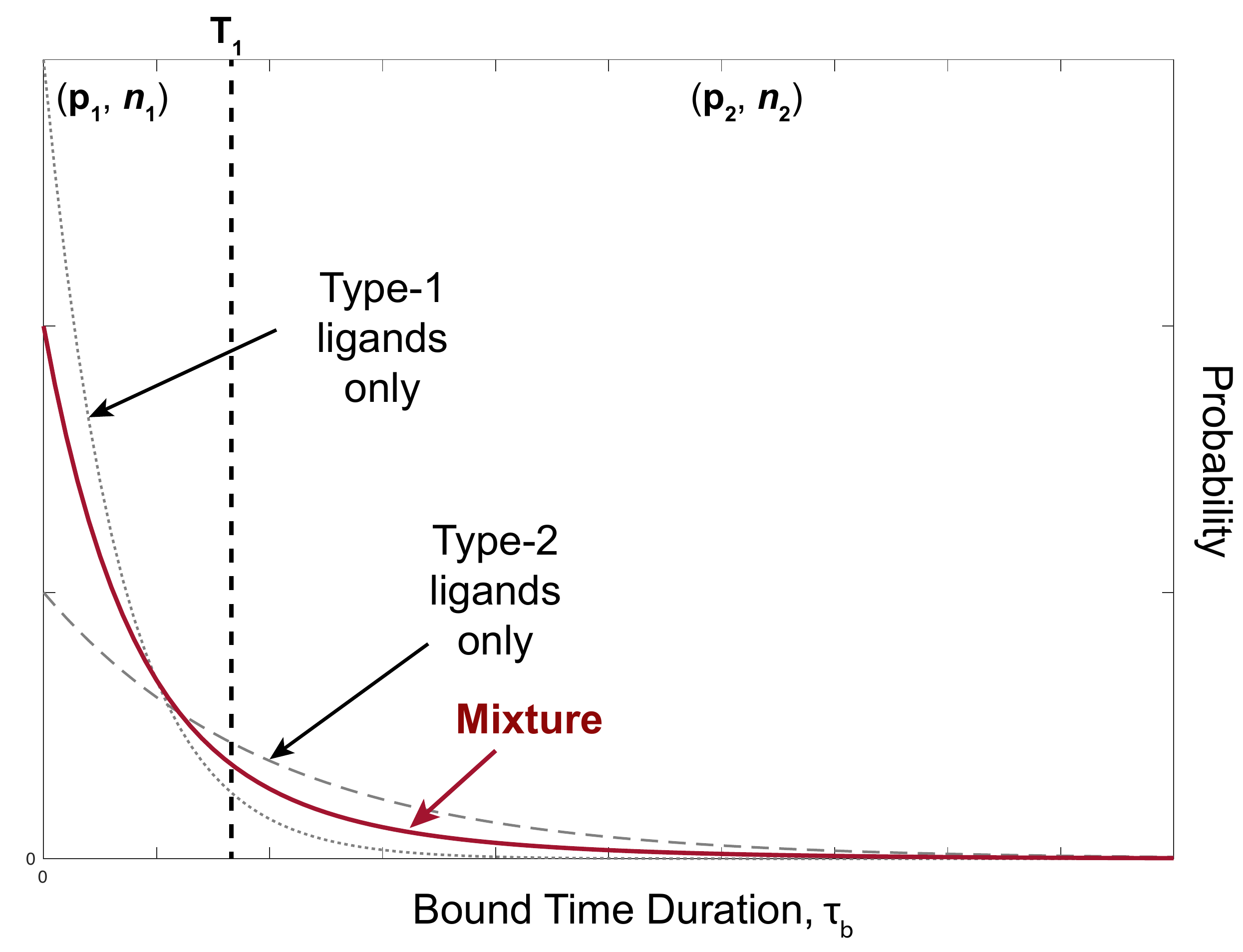}
	\caption{Probability distribution of bound time durations for two ligands and mixture of these ligands. The distribution is separated into two regions by a time threshold ($T_1$). The number of binding events and the probability of observing a binding time duration are given for the corresponding regions.}
	\label{fig:suboptimalbound}
\end{figure}

To address the computational complexity of the optimal estimation scheme for bio-nano devices, Kuscu et al. proposed an alternative method for concentration ratio estimation based on the Method of Moments (MoM) \cite{kuscu2019channel} . This method involves counting the number of receptor binding events with bound time durations that fall within specific time intervals, which are demarcated by time thresholds determined by using the inverse of the ligands' unbinding rates. In the case of two types of ligands that can bind to receptors, there is only one time threshold value as demonstrated in Fig. \ref{fig:suboptimalbound}, which is set by the unbinding rate of the lower affinity ligand, $k^-_1$, i.e., 
\begin{equation}
    \label{eq:times}
    T_1 = v / k^-_1.
\end{equation} 
where $v$ is called the proportionality constant, which can be optimized for improved estimation performance, and type-1 ligand is the lower affinity ligand. The probability of a binding event to have a duration that falls into a time interval between two time thresholds can be obtained as
\begin{align} \label{eq:p_l}
	\p_l &= \int_{T_{l-1}}^{T_l} \p(\tau_b^\prime) d\tau_b^\prime = \sum_{i=1}^2 \alpha_i (e^{- k^-_i T_{l-1}  } - e^{- k^-_i T_l  }) \\ 
	&=\alpha (e^{- k^-_1 T_{l-1}  } - e^{- k^-_1 T_{l}  }) + (1-\alpha) (e^{- k^-_2 T_{l-1}  } - e^{- k^-_2 T_{l}  }), 	\nonumber
\end{align} 
where we set $T_0 = 0$ and $T_2 = \infty$. In matrix notation, \eqref{eq:p_l} can be written as follows
\begin{equation}
    \boldsymbol{\p} =  \boldsymbol{S} \boldsymbol{\alpha},
\end{equation} 
where $\boldsymbol{\p}$ is a $(2\times1)$ probability vector with elements $\p_l$, $ \boldsymbol{\alpha}$ is the $(2\times1)$ vector of ligand concentration ratios, i.e., $[\alpha,1-\alpha]$,  and $\boldsymbol{S}$ is an $(2\times2)$ matrix given by
\begin{equation}
    \boldsymbol{S} = \begin{pmatrix}
    \centering
    e^{- k^-_1 T_{0}  } - e^{- k^-_1 T_{1}  } & e^{- k^-_2 T_{0}  } - e^{- k^-_1 T_{1}  }\\
    e^{- k^-_1 T_{1}  } - e^{- k^-_2 T_{2}  } & e^{- k^-_2 T_{1}  } - e^{- k^-_2 T_{2}  }
\end{pmatrix}.
\end{equation}

Assuming that binding events are independent from each other, the number of binding events with bound time durations that fall within specific time intervals follows a binomial distribution, with the mean and variance given as follows 
\begin{align}
   \nonumber
    \boldsymbol{\E}[\boldsymbol{n}] &= \boldsymbol{\p} N_R, \\
       \boldsymbol{\Var}[\boldsymbol{n}]&= (\boldsymbol{\p} \odot (1-\boldsymbol{\p})) N_R,
\end{align}
where $\boldsymbol{n}$ is a $(2\times1)$ vector with elements $n_i$ which is the number of binding events whose durations fall into the  $i^\text{th}$ time interval between $T_{i-1}$ and $T_i$, and $\odot$ denotes the element-wise product. Using MoM, we can now estimate ligand concentration ratios by comparing the expected number of binding events with durations that fall within specific time interval to the observed number of binding events in the same interval. In other words, we use the first moment to match the predicted and actual number of binding events for each time interval, i.e., 
\begin{align}
\nonumber
    \boldsymbol{n} &= \hat{\boldsymbol{\p}} N_R = \boldsymbol{S} \hat{\boldsymbol{\alpha}} N_R,
    \\
    \hat{\boldsymbol{\alpha}} &= \frac{1}{N_R}  \boldsymbol{S}^{-1} \boldsymbol{n}  = \frac{1}{N_R}  \boldsymbol{W} \boldsymbol{n}, 
\end{align} 
where hat indicates the estimated parameters, and $\boldsymbol{W} = \boldsymbol{S^{-1}}$ is a $(2\times2)$ matrix with elements denoted by $\omega_{i,j}$. The estimated concentration ratios of type-1 and type-2 ligands are then given by 
\begin{equation}
    \hat{\alpha}_l = \left(\frac{1}{N_R}\right) \sum_{i=1}^{2} n_i \omega_{l,i} = \left(\frac{1}{N_R}\right) n_1 \omega_{l,1}+n_2 \omega_{l,2},
\end{equation} 
where $l \in \{1,2\}$. The mean and the variance of the ratio estimator can then be obtained as
\begin{align}
\nonumber
\label{eq:mean_var_subop}
    \boldsymbol{\E}[\boldsymbol{\hat{\alpha}}] &= \frac{1}{N_R} \boldsymbol{W} \mathbf{E}[\boldsymbol{n}] = \boldsymbol{W}\mathbf{p} = \boldsymbol{S^{-1}}\mathbf{p} = \boldsymbol{\alpha} \\
    \Var[\hat{\alpha}_l] &= \frac{1}{N_R^2} \sum_{i=1}^{2} \sum_{j=1}^{2} \omega_{l,i} \omega_{l,j} \Cov[n_i,n_j],
\end{align}
where the covariance function is given as follows
\begin{equation}
    \mathrm{Cov}[n_i,n_j] = \left\{
	\begin{array}{ll}
		\Var[n_i],  & \mbox{if } i = j, \\
		-\p_i \p_j N_R, & \mbox{otherwise.} 
	\end{array}
\right.
\end{equation}
with $\Var[n_i]$ being the $i^\text{th}$ element of the vector $\boldsymbol{\Var}[\boldsymbol{n}]$.

\subsection{Estimation of Ligand Concentrations}
\label{sec:csk_estimation}
If a single type of ligand is used in the modulation scheme, e.g., CSK, ligand concentration can be estimated from the number of bound receptors sampled at equilibrium of the ligand-receptor binding interaction. 

Under equilibrium conditions, the state of a single receptor can be represented with Bernoulli distribution with the success probability $\p_B$ given in \eqref{eq:success_prob_csk} as the probability of success. For $N_R$ receptors, the number of bound receptors $n_B$ can be represented by a binomial distribution i.e., $n_B \sim \mathcal{B}(\p_B,N_R)$, with the mean and variance given by 
\begin{align}
\label{eq:mean_var_c}
\nonumber
    \mu_{n_B} &= \p_BN_R \\
   \sigma^2_{n_B}  &= \p_B (1-\p_B)N_R.
\end{align}
The unbiased estimator of $\p_B$, i.e., $\hat{\p}_B$, is given by
\begin{equation}
    \hat{\p}_B = \frac{n_B}{N_R}.
\end{equation}
By inverting the input-output relation between $c$ and $\p_B$ in \eqref{eq:success_prob_csk}, estimator for the ligand concentration, i.e., $\hat{c}$, can be written as follows
\begin{equation}
    \hat{c} = K_D\frac{\hat{\p}_B}{1-\hat{\p}_B}.
\end{equation}

\section{MC System Model}
\label{sec:sys_model}
We study an MC system with a single pair of MC transmitter and receiver, where the transmitter employs RSK modulation by using two distinct ligand types (type-1 and type-2). The similarity between the two ligand types is quantified by a parameter $\gamma$, defined as the ratio of their unbinding rates, i.e., $\gamma = k_1^-/k_2^-$. The information is encoded in the concentration ratio of the first ligand type, i.e., $\alpha \in [0,1]$, where the subscript is omitted for ease of notation. To decode the transmitted symbol, the receiver estimates the concentration ratio from the bound time statistics of the resulting ligand-receptor interactions on its surface. As a benchmark for evaluating the performance of RSK modulation in this setting, we also consider CSK modulation, in which the information is encoded in the concentration of a single type of ligands, i.e., type-1 ligand. The following assumptions are made for the considered MC system: 
\begin{itemize}
    \item The transmitter releases molecules to the channel as an impulse, i.e., $x(t) = N_{tx} \delta(t)$, where $x(t)$ is the number of transmitted molecules, and $N_{tx}$ is the number of molecules to be transmitted. Note that in the case of RSK, the transmitter releases two types of molecules at the same time instant. The transmitted molecules then propagate in the channel via free diffusion.

  \item Intersymbol interference (ISI) is neglected on the grounds that the signaling interval length $T_s$ is sufficiently large, or there are auxiliary enzymes in the channel that degrade the information molecules \cite{noel2014improving}. 
  
    \item The binding rates of all ligand types are equal to each other. All ligands and receptors are assumed to be monovalent, i.e. a ligand can bind to only one receptor at a time, and vice versa.  
    \item The receiver employs only a single type of receptors on its surface, and all copies of the receptors are independent of each other. Each independent receptor is exposed to the same ligand concentration.
    \item Due to the low-pass characteristics of the MC channel, during the sampling of bound time intervals or the number of bound receptors, the ligand concentrations in the vicinity of the receptors are assumed to be stationary. Variations in the concentration of ligands due to the binding reactions are also assumed to be negligible.
    \item The receiver is assumed to know the unbinding rates of the ligand types used for modulation.

\end{itemize}

Based on this system model, we first investigate the MC channel capacity with RSK in the next section.      

\section{MC Channel Capacity with RSK}
\label{sec:channel_capacity}
Channel capacity is the maximum rate at which information transfer can be transfered reliably through a communication channel and is equal to the mutual information maximized over all input distributions. The input distribution that achieves the channel capacity is called the optimal input distribution and is denoted by $\PP^*(x)$. 

Under regularity conditions, which are discussed in detail in \cite{clarke1994jeffreys,walker1969asymptotic}, the optimal input distribution $\PP^*(x)$ converges asymptotically to the Jeffreys Prior, $\PP_{jp}^*(x)$, as the number of independent receivers, which corresponds to the number of independent receptors in the context of our study, increases\cite{bernardo1979reference,clarke1994jeffreys,berger2009formal}. It has also been shown that $\PP_{jp}^*(x)$ is proportional to the square root of the determinant of the Fisher information matrix \cite{jeffreys1946invariant}, indicating a direct link between information and estimation theories. By combining these two results, for one-dimensional inputs, the optimal input distribution asymptotically becomes proportional to the square root of the scalar Fisher information \cite{jetka2018information,komorowski2019limited}:
\begin{equation}
        {\PP}^{*}(x) \propto \sqrt{
        I(x)},
\label{eq:opt_prob_dist}        
\end{equation} resulting in the approximate channel capacity as follows
\begin{equation}
\label{eq:asmyptotic_capacity_primitive}
    C_A^* = \log_2\bigg((2 \pi e)^{-\frac{1}{2}} \int_\mathcal{X} {\sqrt{ I(x)} dx} \bigg), 
\end{equation} where $\mathcal{X}$ is the one-dimensional input symbol space.
\subsection{MC Channel Capacity with RSK Modulation}
\label{sec:rsk}
Here we derive the capacity of an point-to-point MC channel where the transmitter employs RSK modulation, and the receiver estimates the ligand concentration ratio in its vicinity from the ligand-receptor binding statistics on its surface in order to decode the transmitted symbol. We consider both cases where the receiver employs optimal and suboptimal ratio estimation.

\subsubsection{MC Channel Capacity with RSK Modulation and Optimal Ratio Estimation}
The ratio of the received ligand concentrations can be estimated in an ML manner by maximizing the likelihood of observing a set of bound time intervals $\{\tau_b\}$ over the input space, which, in this case, corresponds to the concentration ratio of ligands, i.e., $\alpha \in [0, 1]$. The log-likelihood of observing $\{\tau_b\}$ given the concentration ratio of type-1 ligands can be written by transforming \eqref{eq:likelihood_boundtimedurations} as follows 
\begin{equation}
\mathcal{L}(\{\tau_b\}|\alpha) = \sum_{i=1}^{N_R} \ln \biggl( k_2^- e^{- k_2^- \tau_{b,i}} \left(1 - \alpha + \alpha \gamma e^{(1 - \gamma) k_2^- \tau_{b,i})} \right) \biggr).    
\end{equation}

The Fisher information can then be derived from the log-likelihood function as follows
\begin{align}\label{fisher_alpha}
I_{RSK}(\alpha) &= -\E \bigg[ \frac{\partial^2}{\partial \alpha^2} \mathcal{L} \left(\{\tau_b \} | \alpha\right) \bigg] \\ \nonumber
&= N_R k_2^- \int_{0}^{\infty} \frac{\left(-1 + \gamma e^{(1 - \gamma) k_2^- \tau_{b} }\right)^2}{1 - \alpha + \alpha \gamma e^{(1 - \gamma) k_2^- \tau_{b}} } e^{-k_2^- \tau_b} d \tau_b.
\end{align} 
By plugging this expression into \eqref{eq:opt_prob_dist} and \eqref{eq:asmyptotic_capacity_primitive}, the optimal input distribution and the approximate channel capacity $C_{RSK}$ can be obtained, respectively, as follows
\begin{equation}
        {\PP}^{*}_{RSK}(\alpha) \propto \sqrt{
        I_{RSK}(\alpha)},
\label{eq:opt_prob_dist_RSK}   
\end{equation}
\begin{equation} 
 C_{RSK} = \log_2 \bigg((2 \pi e)^{-\frac{1}{2}} \int_{0}^{1} {\sqrt{ I_{RSK}(\alpha)} d\alpha}\bigg).
\end{equation}

\subsubsection{MC Channel Capacity with RSK Modulation and Suboptimal Ratio Estimation}
In Section \ref{sec:suboptimal_est}, a suboptimal concentration ratio estimation scheme is introduced to estimate the ratio based on predetermined time intervals that separate binding events according to their bound time durations. Since RSK relies on only two different ligand types, the proposed scheme will have only one time threshold $T$, which distinguishes long binding events from short binding events. Given the concentration ratio of type-1 ligands, $\alpha$, the probability of observing a binding event with a bound time duration longer than $T$ can be written as 
\begin{align}
\nonumber
    \p_T \equiv \p(\tau_b \geq{} T | \alpha) &= \alpha e^{-k_1^- T} + (1-\alpha) e^{-k_2^-T} \\  
    &= e^{-k_2^-T}\left(\alpha e^{(1-\gamma)k_2^-T} + 1- \alpha\right) 
\label{p_t_alpha}
\end{align}
Assuming that only a single binding event is sampled from each independent receptor, the number of binding events that satisfy $\tau_b > T$ is given by a binomial distribution.
\begin{align}
\label{binomial_suboptimal}
    n_T \sim \mathcal{B}(\p_T,N_R).
\end{align}
Fisher information for the suboptimal concentration ratio estimator can then be written as follows
\begin{align}\label{fisher_sub}
\nonumber
&I_{RSK,sub}(\alpha) = -\E \bigg[ \frac{\partial^2}{\partial \alpha^2} \mathcal{L} \left(n_T  | \alpha\right) \bigg] = \left(e^{-\gamma k_2^- T} - e^{-k_2^- T}\right)^2 \\ 
& \times \sum_{n_T=0}^{N_R}  \bigg[\frac{n_T}{\p_T^2} + \frac{N_R-n_T}{(1-\p_T)^2}  \bigg]  \binom{N_R}{n_T} \p_T^{n_T} \left(1-\p_T\right)^{N_R-n_T},
\end{align} 
where the subscript $sub$ indicates the suboptimality of the estimation scheme employed by the receiver.

Finally, $I_{RSK,sub}(\alpha)$ can be plugged into \eqref{eq:opt_prob_dist} and \eqref{eq:asmyptotic_capacity_primitive} to obtain the optimal input distribution and the approximate capacity of the MC channel with the suboptimal ratio estimator as follows
\begin{equation}
        {\PP}^{*}_{RSK,sub}(\alpha) \propto \sqrt{
        I_{RSK,sub}(\alpha)},
\label{eq:opt_prob_dist_RSK_sub}   
\end{equation}
\begin{equation} 
 C_{RSK,sub} = \log_2 \bigg({(2 \pi e)^{-\frac{1}{2} } \int_{0}^{1} {\sqrt{ I_{RSK,sub}(\alpha)} d\alpha}}\bigg).
\end{equation}

\subsection{MC Channel Capacity with CSK Modulation}
\label{sec:CSK_capacity}
We also investigate the approximate capacity of an MC channel with CSK modulation as a benchmark for an in-depth evaluation of the RSK performance. In CSK, information is encoded in the concentration of a particular type of ligands, and the detection is performed by sampling the number of bound receptors in each signaling interval at a pre-defined sampling time, which is typically taken as the peak time of the ligand concentration in the vicinity of the receiver. The Fisher information in this case can be calculated as follows:
\begin{align}
\label{eq:fisher_CSK}
\nonumber
I_{CSK}(c) =& -\E \bigg[ \frac{\partial^2}{\partial c^2} \mathcal{L} \left(n_B | c \right) \bigg] \\ 
    =&-\sum_{n_B=0}^{N_R} \Bigg( \frac{\partial^2 \p_B}{\partial c^2}  \bigg[\frac{n_B}{\p_B} - \frac{N_R-n_B}{1-\p_B} \bigg] \\ \nonumber
    &- \bigg(\frac{\partial \p_B}{\partial c} \bigg)^2 \bigg[\frac{n_B}{\p_B^2} + \frac{N_R-n_B}{(1-\p_B)^2} \bigg] \Bigg) \\ \nonumber
    &\times \binom{N_R}{n_B} \p_B^{n_B} (1-\p_B)^{N_R-n_B}.
\end{align}

In an MC system with a power-limited transmitter, the input symbol space in \eqref{eq:asmyptotic_capacity_primitive} is limited by the maximum concentration of ligands that the transmitter can release into the channel. In this case, the approximate capacity of the channel can be obtained as
\begin{equation} 
 C_{CSK} = \log_2 \Bigg({(2 \pi e)^{-\frac{1}{2} } \int_{0}^{c_{Rx,max}} {\sqrt{ I_{CSK}(c) } dc\Bigg)}},
\end{equation} 
where $c_{Rx,max}$ is the ligand concentration in the vicinity of the receptors at the sampling time, which corresponds to the maximum ligand concentration that can be transmitted by the transmitter, scaled by the CIR of the diffusion-based MC channel. Finally, the optimal input distribution can be obtained by plugging \eqref{eq:fisher_CSK} into \eqref{eq:opt_prob_dist} as follows:
\begin{equation}
    {\PP}^{*}_{CSK}(c) \propto \sqrt{
    I_{CSK}(c)}.
\label{eq:opt_prob_dist_CSK}   
\end{equation}

\subsection{Information Theoretical Analysis}
\label{sec:capacity_result}
We numerically evaluate the approximate capacity of the end-to-end MC channel with RSK and CSK modulation under varying system settings. We analyze the channel capacity with RSK with respect to the level of similarity between the two ligand types and the number of receptors, whereas the channel capacity with CSK is analyzed as a function of the maximum received concentration and the number of receptors. We also present the corresponding optimal input distributions for the RSK and CSK scenarios. Default values of the number of receptors and the similarity parameter used in the analyses are $N_R = 1000$ and $\gamma = 5$, respectively.

\subsubsection{\textbf{Optimal Input Distribution}}
\begin{figure*}
	\centering
	\begin{subfigure}[b]{0.32\textwidth}
		\centering
		\includegraphics[width=\textwidth]{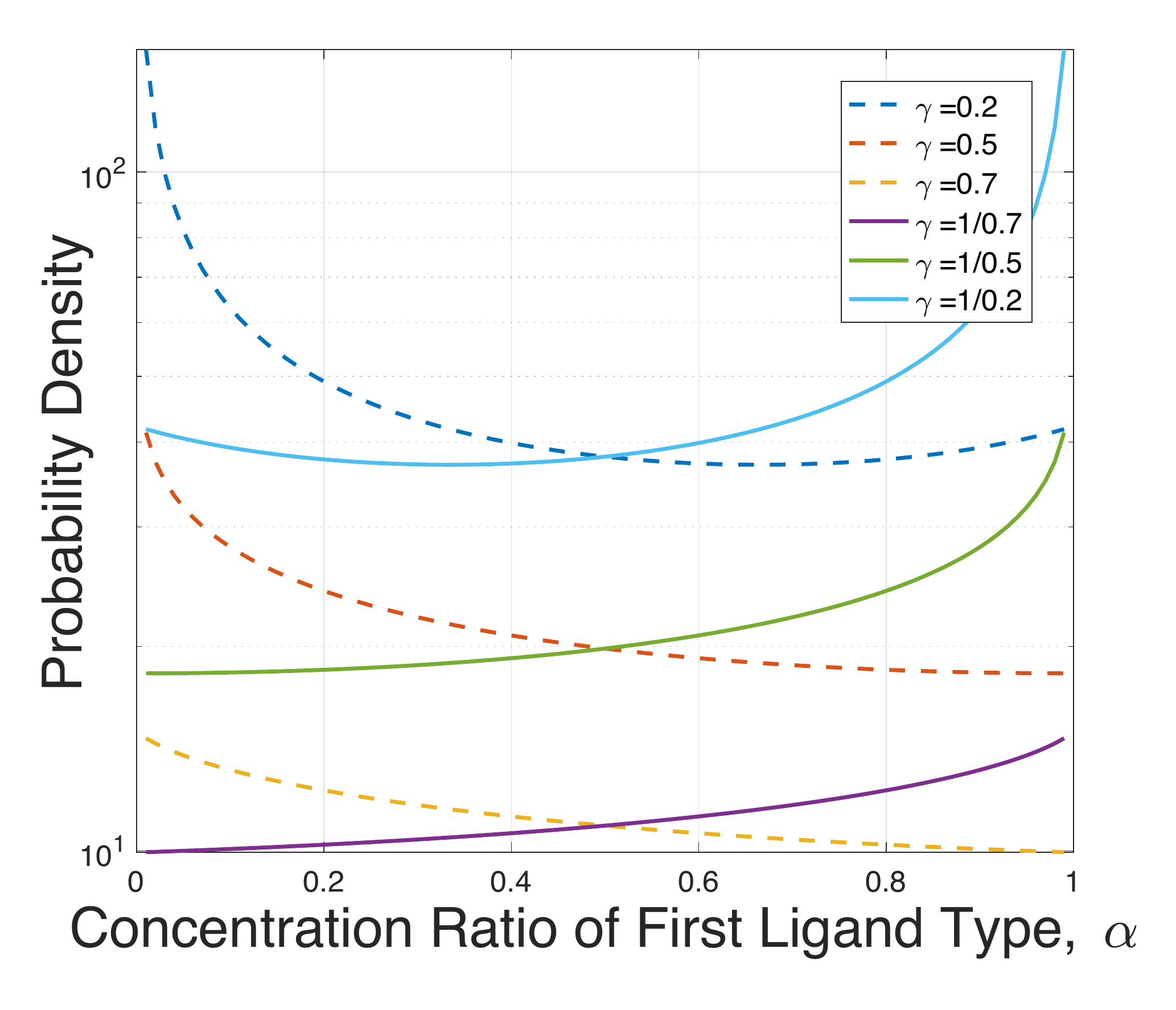}
		\caption{}
		\label{fig:OID_RSK_optimal}
	\end{subfigure}
	\hfill
	\begin{subfigure}[b]{0.32\textwidth}
		\centering
		\includegraphics[width=\textwidth]{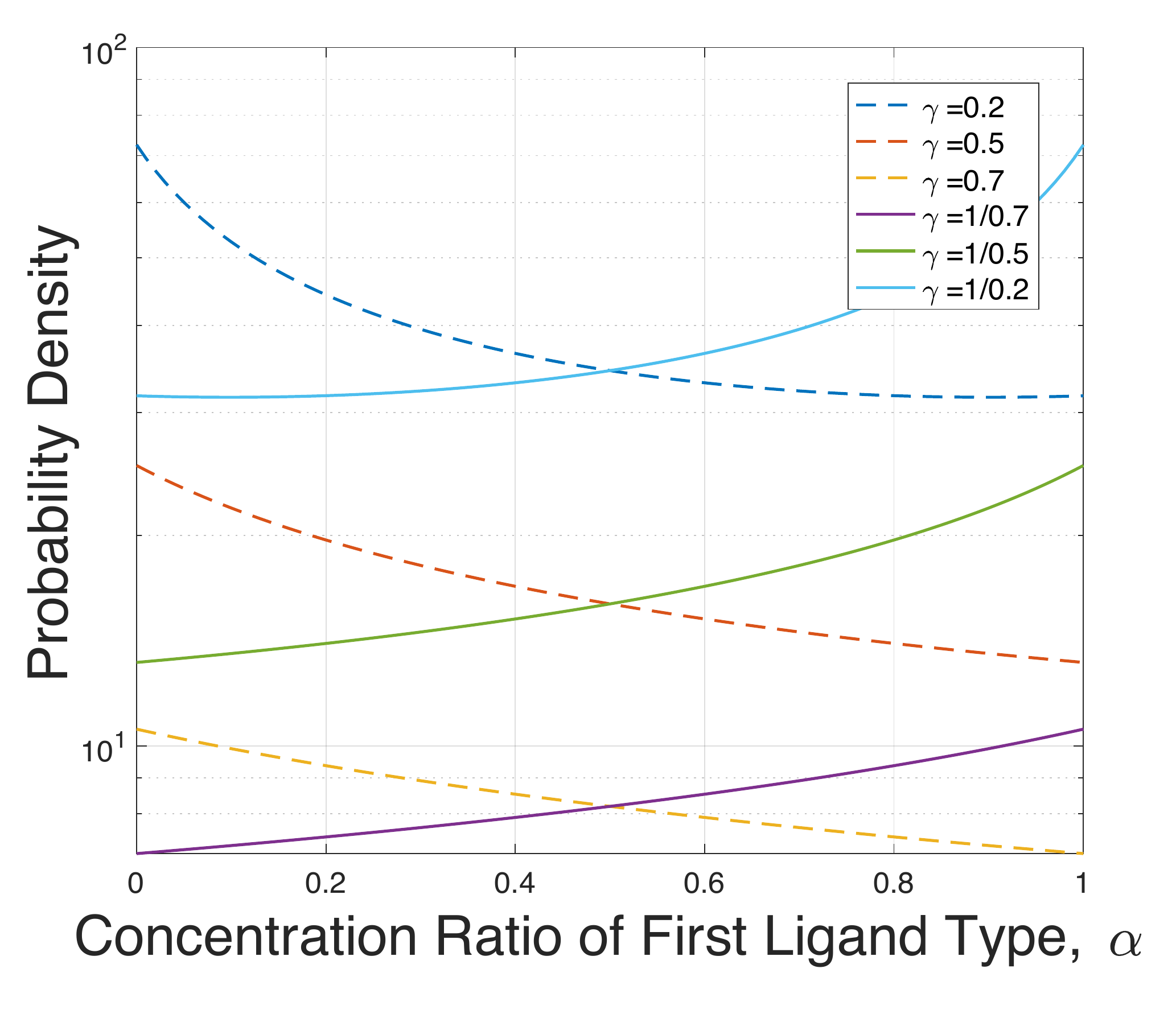}
		\caption{}
		\label{fig:OID_RSK_suboptimal}
	\end{subfigure}
	\hfill
	\begin{subfigure}[b]{0.32\textwidth}
		\centering
		\includegraphics[width=\textwidth]{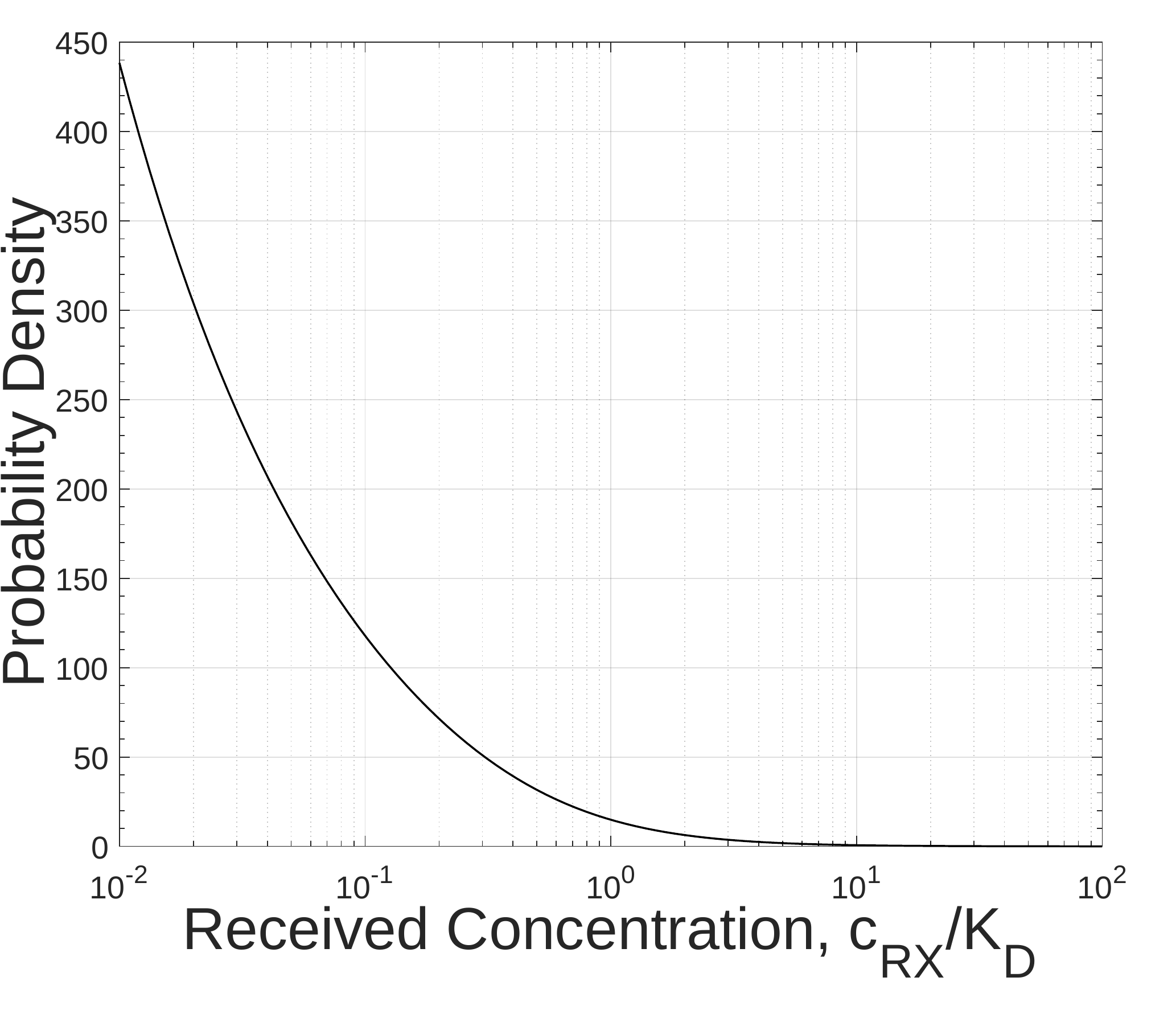}
		\caption{}
		\label{fig:OID_CSK}
	\end{subfigure}
	\caption{Optimal input distribution for end-to-end MC channel (a) with RSK modulation and optimal ratio estimator, (b) with RSK modulation and suboptimal ratio estimator, (c) with CSK modulation.}
\end{figure*}

The optimal input distribution that achieves the capacity for the MC channel with RSK modulation is given in Figs. \ref{fig:OID_RSK_optimal} and \ref{fig:OID_RSK_suboptimal}, for cases where the receiver utilizes optimal and suboptimal concentration ratio estimators, respectively. Here, we take the input as the ratio of the concentration of the first ligand type to the total ligand concentration, i.e., $\alpha \in [0,1]$, and the optimal input distributions are shown for different values of the similarity parameter, $\gamma = k_1^-/k_2^-$. 

Our first observation is that the optimal input distribution is symmetric for $\gamma = a$ and $\gamma = 1/a$, with $a \in (0,1)$. Additionally, the optimal input distribution favors the ligand type with lower affinity for the receptors, i.e., the one with higher unbinding rate. This can be explained by the fact that it is less likely to find a receptor bound to a ligand with higher unbinding rate at equilibrium, decreasing the number of samples informative of this particular ligand type. As a result, the optimal input distribution shifts the concentration ratio towards the less likely binding ligand, increasing their proportion among all the bound ligands at equilibrium. This preference becomes more prominent as the similarity between the ligand types decreases. The same trend can be observed in both the optimal and suboptimal cases, although the preference for the less cognate ligand is more prominent in the optimal case. 

The optimal input distribution for MC channel with CSK is given in Fig. \ref{fig:OID_CSK}. The numerical analysis for the calculation of the optimal input distribution does not have a constraint on the maximum received concentration $c_{Rx,max}$, where the input is the ligand concentration in the vicinity of the receiver. We observe that the optimal input distribution favors low ligand concentrations, which is consistent with the results of Einolghozati et al. in \cite{einolghozati2011capacity}.

\subsubsection{\textbf{End-to-End Channel Capacity}}
\begin{figure*}
	\centering
	\begin{subfigure}[b]{0.32\textwidth}
		\centering
		\includegraphics[width=\textwidth]{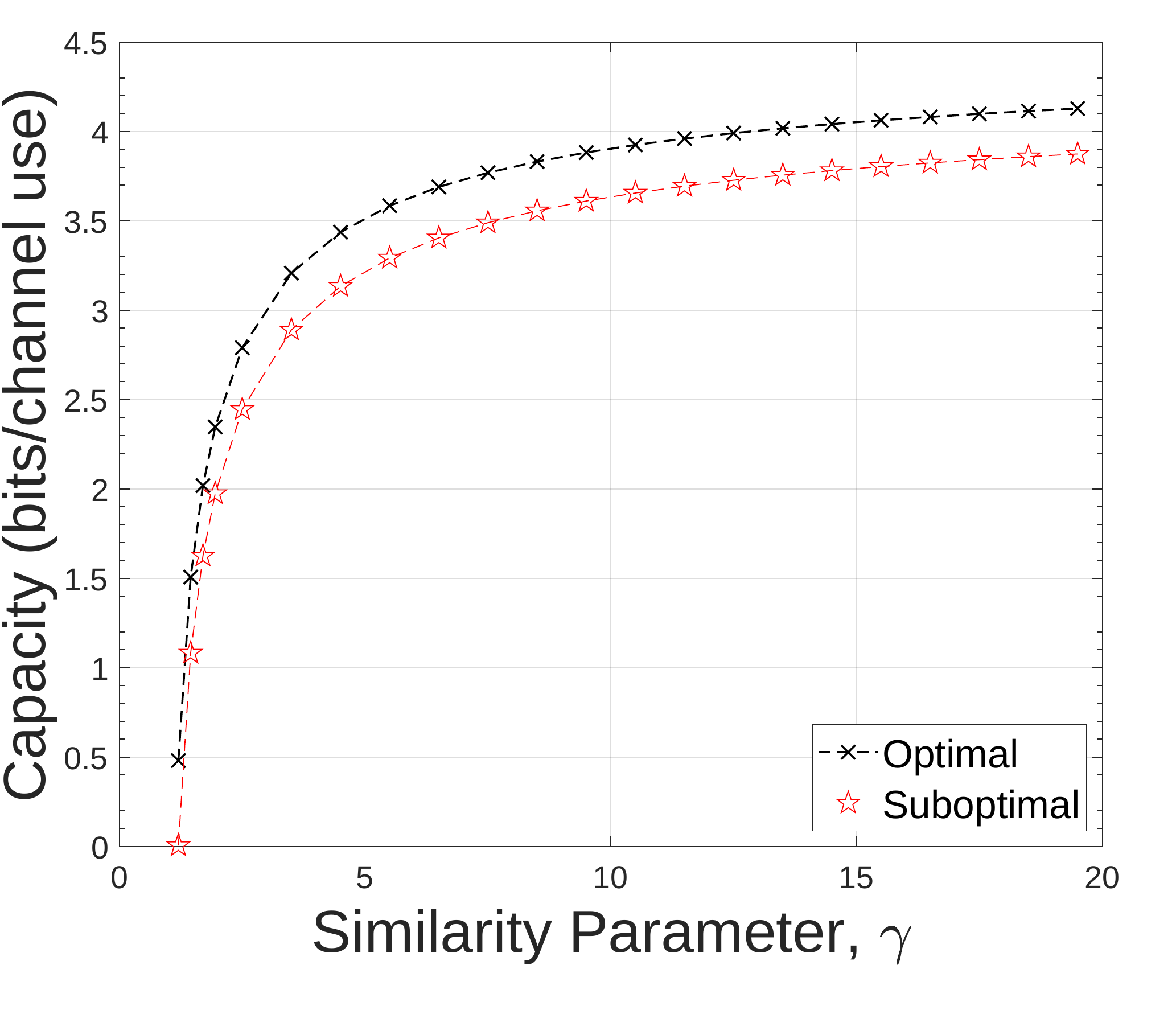}
		\caption{}
		\label{fig:capacity_gamma}
	\end{subfigure}
	\hfill
	\begin{subfigure}[b]{0.32\textwidth}
		\centering
		\includegraphics[width=\textwidth]{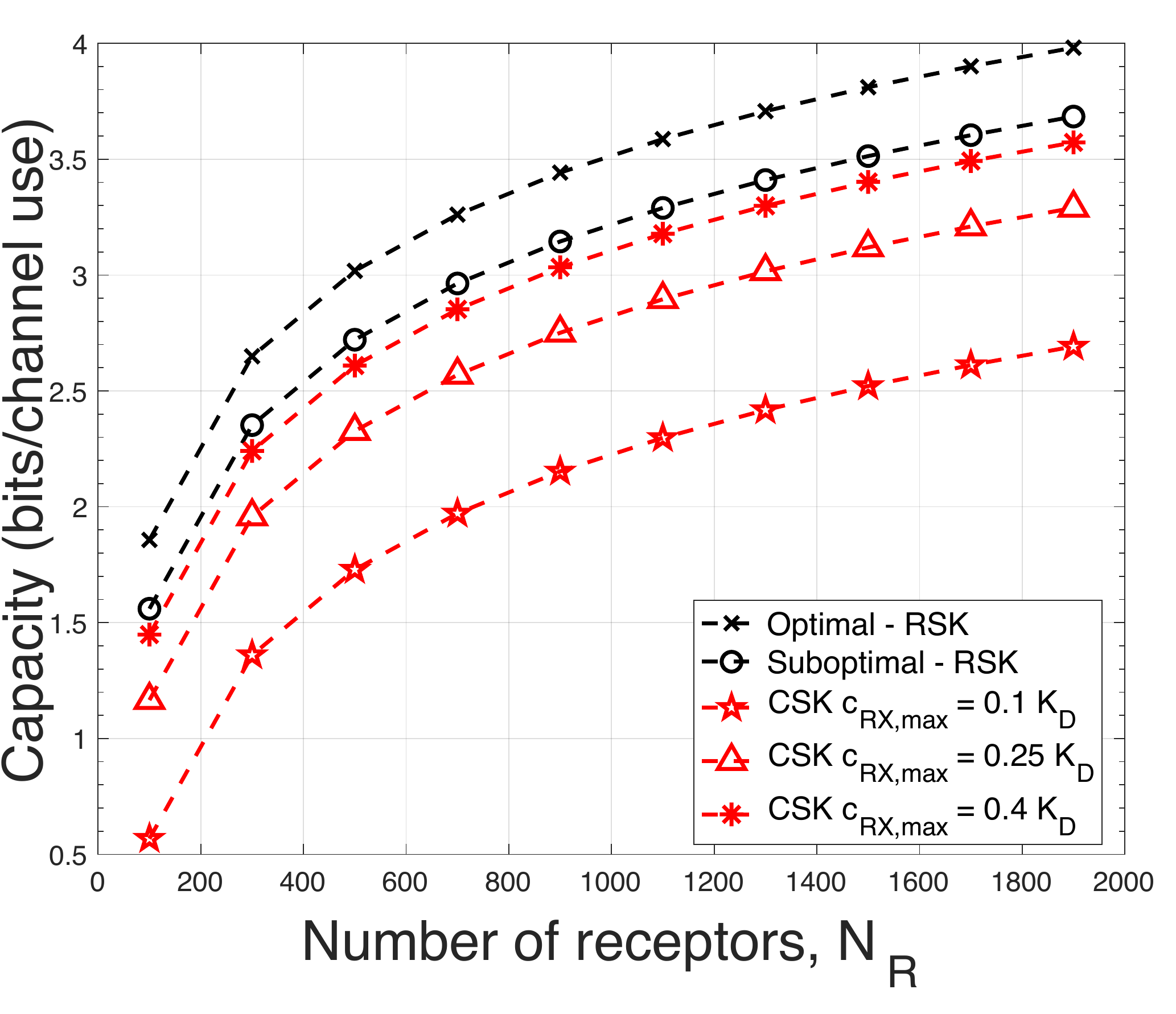}
		\caption{}
		\label{fig:capacity_N}
	\end{subfigure}
	\hfill
	\begin{subfigure}[b]{0.32\textwidth}
		\centering
		\includegraphics[width=\textwidth]{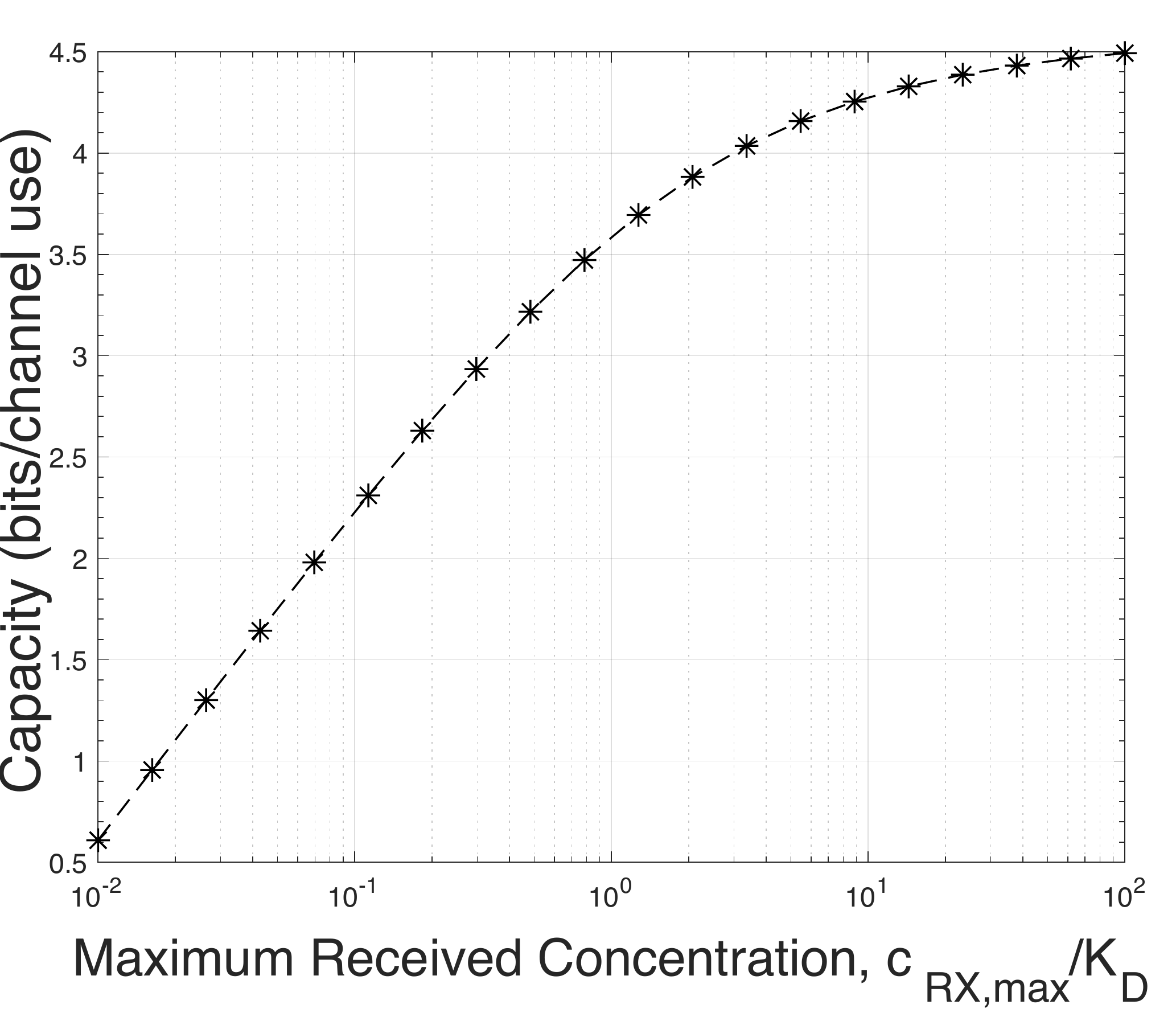}
		\caption{}
		\label{fig:capacity_CSK}
	\end{subfigure}
	\caption{End-to-end MC channel capacity (a) with RSK modulation as a function of similarity parameter $\gamma$, (b) with RSK and CSK modulation as a function of number of receptors $N_R$, (c) with CSK modulation as a function of maximum received concentration $c_{Rx,max}$.}
\end{figure*}

We first analyze the impact of the similarity between the ligand types, quantified by $\gamma = k_1^-/k_2^-$, on the capacity of the MC channel with RSK. The results are provided in Fig. \ref{fig:capacity_gamma} for cases where the receiver estimates the ligand concentration ratio either optimally or suboptimally. As $\gamma$ increases, the similarity between the ligand types in terms of their unbinding rates from the receptors decreases, increasing their distinguishability by the receiver. As is seen in Fig. \ref{fig:capacity_gamma}, this leads to higher channel capacities, which saturate around $4$ bits/channel use. When $\gamma \approx 1$, the ligand types are hardly distinguishable, resulting in a channel capacity close to $0$. However, even a small difference in the unbinding characteristics of the ligands significantly improves the channel capacity. More importantly, the channel capacity obtained with the suboptimal estimator is close to that obtained with the optimal estimator, while having much lower complexity. This makes RSK modulation feasible for resource-constrained bio-nano devices, as the suboptimal estimator can be implemented with a simple single-threshold kinetic proofreading (KPR) scheme, similar to those already used in living cells \cite{mckeithan1995kinetic}. 

Our next analysis focuses on the impact of the number of receptors, $N_R$, on the channel capacity. As the receptors are considered to be independent of each other, $N_R$ determines the number of independent samples taken from the receptors in each signaling interval for estimating the ligand concentration ratio in the case of RSK, and the ligand concentration in the case of CSK. In Fig. \ref{fig:capacity_N}, we compare the performance of RSK and CSK modulatiosn under power-limited conditions, where the transmitter has an upper bound on the number of molecules it can transmit. We see that increasing the number of independent samples, i.e., $N_R$, has a significant effect on the capacity, as it improves the accuracy of the estimation performed by the receiver. The power limitation of the transmitter is translated into an upper-bound on the ligand concentration in the vicinity of receptors, as the free diffusion channel is deterministic in terms of molecule concentration. Accordingly the maximum received concentration is set to $c_{Rx,max} = 0.1~K_D$, $c_{Rx,max} = 0.25~K_D$ and $c_{Rx,max} = 0.4~K_D$ for CSK modulation. We do not consider the effect of power limitation on the RSK performance, as the concentration ratio is invariant to the total number of molecules released by the transmitter. These results align with our discussion in the Introduction, showing that RSK becomes advantageous in terms of channel capacity when the received concentration is upper-bounded. This advantage of RSK over CSK can be particularly significant when the transmitter has a limited molecule reservoir or relies on fluctuating molecule harvesting or production processes. Additionally, RSK may have advantage over CSK in mobile MC scenarios, which we will discuss in Section \ref{sec:mobil}. In these cases, the mobility of the transmitter and/or the receiver can result in a time-varying CIR and received concentration profile, which can degrade the performance of the MC system.

For the completeness of the analysis, we also provide the approximate channel capacity for CSK as a function of maximum received concentration $c_{Rx,max}$ in Fig. \ref{fig:capacity_CSK}. The results show that the asymptotic channel capacity obtained when the $c_{Rx,max}$ is much larger than the dissociation constant of the ligand-receptor pair $K_D$ saturates around $4.5$ bits/channel use, which is only slightly higher than the asymptotic channel capacity of the RSK obtained when the similarity between the utilized ligand types is low (see Fig. \ref{fig:capacity_gamma}). 

The results of the information theoretical analysis demonstrate that the RSK modulation may be able to address some of the limitations of the CSK modulation without sacrificing channel capacity. This is further investigated in the error performance analysis, where we consider a mobile MC scenario with a time-varying CIR. 

\section{Error Performance of RSK in Mobile MC}
\label{sec:mobil}
In this section, we analyze the error performances of Quadrature-RSK (Q-RSK) and Quadrature-CSK (Q-CSK) modulations in a mobile MC scenario. In this scenario, the transmitter and receiver undergo random walk, which is a widely used model for the random movement of living cells and passive micro/nano robots \cite{ahmadzadeh2018stochastic}. We assume that all the symbols have equal probability of being transmitted for both Q-RSK and Q-CSK, and the receiver utilizes the suboptimal estimation of concentration ratio for Q-RSK.

We assume that the transmitter and receiver have equal diffusion coefficients that are much lower than the diffusion coefficient of the information molecules (i.e., ligands). Time-varying position of the transmitter, i.e., $X_{tx}(t),Y_{tx}(t),Z_{tx}(t)$, and the receiver, i.e., $X_{rx}(t),Y_{rx}(t),Z_{rx}(t)$, undergoing random walk can be modeled in Cartesian coordinates as follows
\begin{align}
\label{eq:coordinate_gaussians} 
    \nonumber
   X_{tx}(t) &\sim \mathcal{N}(x_{0,tx},2D_{tx,rx}t),  X_{rx}(t) \sim \mathcal{N}(x_{0,rx},2D_{tx,rx}t),\\ 
   \nonumber Y_{tx}(t) &\sim \mathcal{N}(y_{0,tx},2D_{tx,rx}t), Y_{rx}(t) \sim \mathcal{N}(y_{0,rx},2D_{tx,rx}t),  \\  Z_{tx}(t) &\sim \mathcal{N}(z_{0,tx},2D_{tx,rx}t),  Z_{rx}(t) \sim \mathcal{N}(z_{0,rx},2D_{tx,rx}t),
\end{align} 
where $\mathcal{N}(\mu,\sigma^2)$ denotes Gaussian distribution with mean $\mu$ and variance $\sigma^2$, $(x_{0,tx},y_{0,tx},z_{0,tx})$ corresponds to the initial position coordinates of the transmitter, $(x_{0,rx},y_{0,rx},z_{0,rx})$ are the initial position coordinates of the receiver, $D_{tx,rx}$ is the common diffusion coefficient of the transmitter and receiver. Without loss of generality, we assume that the transmitter is initially located in the origin, i.e., $\Vec{r}_{tx}(t = 0) = [0,0,0]$ and the receiver is at the position $\Vec{r}_{rx}(t = 0) = [x_{0,rx},0,0]$ where $\Vec{r}_{tx}$ and $\Vec{r}_{rx}$ denotes the position vectors of transmitter and receiver, respectively.

 We adopt the notation and formulation introduced in \cite{ahmadzadeh2017statistical} to model the CIR of the mobile MC channel as follows
\begin{equation}
\label{eq:CIR}
    h(t,\tau) = \frac{1}{(4\pi D \tau)^{3/2}}\mathrm{exp}\bigg(- \frac{r(t)^2}{4D\tau}\bigg),
\end{equation} 
where $r(t)$ is the transmitter-receiver distance at time $t$, $\tau$ is the relative time of sampling at the receiver with respect to the time of release of molecules from the transmitter $t = t_R$, such that $\tau = t - t_R$. For a particular signaling interval starting with the transmission time $t = t_R$, CIR can be assumed to be time-independent, since the diffusion coefficients of the transmitter and receiver are much lower than that of the information molecules. Based on this assumption, for a particular signaling interval, CIR becomes $h(\tau)$. We assume that the receiver takes its samples at the peak time of received ligand concentration, i.e., $\tau = \tau_{peak}$.

\subsection{Statistics of Transmitter-Receiver Distance}
With the given initial coordinate choices of the transmitter $\Vec{r}_{tx}(t = 0)$, and the receiver $\Vec{r}_{rx}(t = 0)$, the distributions of the transmitter-receiver distances at each individual coordinate $X_D(t),Y_D(t),Z_D(t)$ become Gaussian distributions given by
\begin{align}
\label{eq:x_y_z_dist}
 \nonumber
    &X_{D}(t) = X_{rx}(t)-X_{tx}(t) \sim \mathcal{N}(x_{0,rx},4D_{tx,rx}t), \\
    \nonumber 
   &Y_{D}(t) = Y_{rx}(t)-Y_{tx}(t) \sim \mathcal{N}(0,4D_{tx,rx}t), \\
   &Z_{D}(t) = Z_{rx}(t)-Z_{tx}(t) \sim \mathcal{N}(0,4D_{tx,rx}t).
\end{align}
Then the transmitter-receiver distance $r(t)$ is given by
\begin{equation}
\label{eq:distance_dist}
    r(t) = \sqrt{X_{D}(t)^2+Y_{D}(t)^2+Z_{D}(t)^2},
\end{equation}
$r(t)$ can be shown to follow a scaled noncentral Chi distribution, and its expected value and variance at time $t$ can be given as
\begin{align}
\label{eq:mean_of_distance_dist}
 \nonumber
 \mu_r(t) &= \sqrt{4D_{tx,rx}t} ~\E[B], \\
   \sigma^2_r(t) &= 4D_{tx,rx}t ~\Var[B],
\end{align}
where
\begin{equation}
    B = \sqrt{\frac{X_{D}(t)^2}{4D_{tx,rx}t}+\frac{Y_{D}(t)^2}{4D_{tx,rx}t}+\frac{Z_{D}(t)^2}{4D_{tx,rx}t}
   }
\end{equation} 
is an auxiliary parameter that follow a noncentral Chi distribution with three degrees of freedom, i.e., $k=3$ and its mean and variance are given by
\begin{align}
    \E[B] &= \sqrt{\frac{\pi}{2}}L_{(1/2)}^{(k/2-1)}\bigg(-\frac{\lambda_n^2}{2}\bigg) = \sqrt{\frac{\pi}{2}}L_{(1/2)}^{(1/2)}\bigg(-\frac{\lambda_n^2}{2}\bigg), \\
    \nonumber
    \Var[B] &= k + \lambda_n^2 - (E[B])^2 = 3 + \lambda_n^2 - (E[B])^2.
\end{align} 
Here $L_{(1/2)}^{(k/2-1)}$ is the Laguerre function, and the noncentrality parameter $\lambda_n$ for $B$ is calculated by the formula
\begin{equation}
     \lambda_n =  \sqrt{\sum_{i=1}^{k=3}\bigg(\frac{\mu_i}{\sigma_i}\bigg)^2}= \sqrt{\frac{x_{0,rx}^2}{4D_{tx,rx}t}}.
\end{equation}

\subsection{Statistics of Peak Time of Received Ligand Concentrations}
Upon the transmitter's release of molecules at time $t = t_R$, the ligand concentration in the receiver's vicinity attains its peak at time $\tau = \tau_{peak}$, which is obtained by solving $\frac{dh(\tau)}{d\tau}=0$ as follows
\begin{equation}
\label{eq:tau_peak}
    \tau_{peak} = \frac{r(t_R)^2}{6D}.
\end{equation} 
Consequently, the random variable $\tau_{peak}$ depends on the random transmitter-receiver distance at the transmission time $t_R$, i.e., $r(t_R)$, and can be expressed as
\begin{equation}
\label{eq:peak_time}
    \tau_{peak}(t_R) = \frac{X_{D}(t_R)^2+Y_{D}(t_R)^2+Z_{D}(t_R)^2}{6D}.
\end{equation}
We know that the sum of squares of three unit-variance Normal distributions follows a noncentral Chi-squared distribution \cite{sankaran1959non}. If we define an auxiliary random variable $A$ as
\begin{equation}
\label{eq:auxillary_nonchisquare}
    A = \frac{X_{D}(t_R)^2+Y_{D}(t_R)^2+Z_{D}(t_R)^2}{4D_{tx,rx}t_R},
\end{equation} 
then $A$ follows a noncentral Chi-squared distribution with three degrees of freedom, i.e., $k=3$, and with noncentrality parameter $\lambda = x_{0,rx}^2/4D_{tx,rx}t_R$. The mean and variance of the auxiliary distribution $A$ can then be calculated as follows
\begin{align}
\nonumber
    \E[A] &= k + \lambda = 3 + \frac{x_{0,rx}^2}{4D_{tx,rx}t_R},\\ 
    \Var[A] &= 2k + 4\lambda = 6 + 4\frac{x_{0,rx}^2}{4D_{tx,rx}t_R}.
\end{align} 
Upon substitution of $X_{D}(t_R)^2+Y_{D}(t_R)^2+Z_{D}(t_R)^2 = 4 D_{tx,rx}t_R A$ in \eqref{eq:peak_time}, we obtain
\begin{equation}
    \tau_{peak}(t_R) = \frac{4D_{tx,rx}t_R}{6D} A,
\end{equation} 
implying that $\tau_{peak}(t_R)$ follows a scaled noncentral Chi-squared distribution. Therefore, the mean and variance of the peak time distribution are
\begin{align}
\nonumber
    \E[\tau_{peak}(t_R)] &= \frac{12D_{tx,rx}t_R + x_{0,rx}^2}{6D}, \\
     \Var[\tau_{peak}(t_R)] &= \frac{24D_{tx,rx}^2t_R^2 + 4 x_{0,rx}^2 D_{tx,rx}t_R}{9D^2}.
\end{align}

\subsection{Statistics of Received Ligand Concentrations}
\label{sec:stat-rec-lig-conc}
The concentration of ligands in the vicinity of the receiver at the sampling time depends on the transmitter-receiver distance at the time of transmission and the sampling time (i.e., peak time of received ligand concentration), both of which are random variables in a mobile MC scenario. 

The nonlinear relationship between the distance and the concentration in the vicinity of the receiver after the transmitter releases its molecules at $t = t_R$ is defined as
\begin{align}
\label{eq:c_received}
\nonumber
    c(t_R + \tau)&= N_{tx,m}h(t_R,\tau)\\ &= N_{tx,m} \frac{1}{(4\pi D \tau)^{3/2}}\mathrm{exp}\bigg(- \frac{r(t_R)^2}{4D\tau}\bigg),
\end{align} 
where $N_{tx,m}$ is the number of transmitted molecules for symbol $m$. When the receiver takes its samples at the peak of the concentration, \eqref{eq:c_received} can be simplified by plugging $\tau = \tau_{peak}$ in \eqref{eq:tau_peak}. The simplified relationship between the distance and the received ligand concentration becomes
\begin{equation}
\label{eq:c_and_dist_relation}
    c(t_R + \tau_{peak} ) = N_{tx,m}\bigg(\frac{2\pi r(t_R)^2}{3}\bigg)^{-3/2} \exp(-3/2).
\end{equation}
Using the expected value and variance of the transmitter-receiver distance $r(t)$ given in \eqref{eq:mean_of_distance_dist}, 
the mean and variance of received ligand concentration $c(t_R + \tau_{peak} )$ can be approximately calculated using the Delta method as follows \cite{casella2021statistical}
\begin{align}
\label{eq:mean_of_concentration}
\nonumber
     \mu_c &= \mathrm{exp}(-3/2) \left(\frac{3}{2\pi}\right)^{3/2}\frac{N_{tx,m}}{\mu_r^3}\left[ 1 + \frac{6\sigma^2_r}{\mu_r^2}    \right],\\
     \sigma^2_c &= \mathrm{exp}(-3)\left(\frac{3}{2\pi}\right)^{3}\frac{9 N_{tx,m}^2 \sigma^2_r}{\mu_r^8} \left[ 1 + \frac{8\sigma^2_r}{\mu_r^2}  \right].
\end{align}

An analogous analysis can be performed for the case where the receiver takes the samples at a fixed time $\tau_s$ instead of the peak time. To this end, we substitute $\tau = \tau_s$ into the CIR equation, resulting in  
\begin{equation}
\label{eq:c_fixed}
    c_f(t_R + \tau_s) =N_{tx,m} \frac{1}{(4\pi D\tau_s)^{3/2}}\mathrm{exp}\left(- \frac{r(t_R)^2}{4D\tau_s}\right).
\end{equation}
The Delta method can then be used to approximate the mean and variance of the received concentration as follows 
 \begin{align}
     \mu_{c_f} = &\frac{N_{tx,m}}{(4 \pi D \tau_s)^{3/2}} \mathrm{exp} \left( \frac{-\mu_r^2}{4D\tau_s}\right) 
     \left[1 + \frac{\sigma^2_r \mu_r^2}{8D^2\tau_s^2} - \frac{\sigma^2_r}{4D\tau_s} \right], \\
     \sigma^2_{c_f} = &\frac{4N_{tx,m}^2 \sigma_r^2}{(4D\tau_s)^5\pi^3}\mathrm{exp}\left(\frac{-2\mu_r^2}{4D\tau_s}\right) 
     \left[ \mu_r^2 + \frac{\sigma_r^2}{2}-\frac{\mu_r^2\sigma_r^2}{2D\tau_s} + \frac{\mu_r^4\sigma^2_r}{8D^2\tau_s^2}
    \right].  \nonumber
 \end{align}

\subsection{Received Signal Statistics}
\label{sec:stat_param}
In MC receivers with ligand receptors, molecular signals in the form of ligand concentration (CSK) or ligand concentration ratio (RSK) are sampled through ligand receptors on the receiver surface. As such, the statistics of ligand-receptor binding interactions in response to received ligand concentration and ligand concentration ratio are utilized for decoding the transmitted messages. In the case of CSK, the number of bound receptors is sampled at the sampling time for decoding. On the other hand, for RSK, as discussed in Section \ref{sec:optimal}, the ligand concentration ratio is estimated through the bound time statistics of the receptors at the sampling time, which is subsequently used for decoding. 

As the time-varying CIR in the mobile case with RSK affects both types of ligands equally, the concentration ratio of ligands in the receiver's vicinity remains unchanged, assuming that ISI is neglected. Thus, the mean and variance of the concentration ratio estimator, given in \eqref{eq:mean_var_subop}, become the first two moments of the received signal statistics for RSK, which can be assumed to be Gaussian distributed when the number of independent receptor bound time samples, which is equal to the number of independent receptors, is sufficiently high. 

However, the received concentration becomes a random variable due to random CIR resulting from mobility, as discussed in Section \ref{sec:stat-rec-lig-conc}. As such, the mean and variance of the number of bound receptors sampled at the sampling time, i.e., $\mu_{n_B}$ and $\sigma^2_{n_B}$, can be calculated using the law of total mean and variance. Assuming that the received ligand concentration at the sampling time follows a Gaussian distribution with mean $\mu_c$ and variance $\sigma_c^2$, the mean and variance of the number of bound receptors can be obtained as follows 
\begin{align}
\label{eq:mean_var_nb}
\nonumber
    \mu_{n_B} &= \E[\E[n_B|c]] \\
    &= \int_{0}^{\infty} \frac{c}{c+K_D}N_R\frac{1}{\sigma_c\sqrt{2\pi}}\exp\bigg(-\frac{1}{2}\bigg(\frac{c-\mu_c}{\sigma_c}\bigg)^2 \bigg) dc, \\ 
    \nonumber
    \sigma^2_{n_B} &= \E[\Var[n_B|c]] + \Var[\E[n_B|c]] \\ \nonumber
        &= \int_{0}^{\infty} \bigg( \frac{cN_R}{(c+K_D)^2}+\bigg[\frac{cN_R}{c+K_D} - \mu_{n_B} \bigg]^2\bigg) \\ 
    &\times \exp\bigg(-\frac{1}{2}\bigg(\frac{c-\mu_c}{\sigma_c}\bigg)^2 \bigg) dc.
\end{align} 

In the case of a large number of independent receptors, the number of bound receptors at the sampling time can be assumed to follow Gaussian distribution, as the receptors are independent and identically distributed.

\subsection{Transmit Signal Design}
Prior to the design of the detection schemes with optimal decision rules for Quadrature-RSK (Q-RSK) and Quadrature-CSK (Q-CSK), we first optimize the set of transmit signals to obtain a quadrature constellation design that maximizes error performance. We use the Chernoff upper bound for pairwise error probability as a design metric in the optimization of transmit signals, which lends itself to analytical expressions when the channel transition probabilities can be approximated as Gaussian. 

Chernoff upper bound for pairwise error probability for two distributions, $\PP(Y|X=x_1)$ and $\PP(Y|X=x_2)$, is given by \cite{fukunaga2013introduction}
\begin{align} \label{eq:chernoff-bound}
 \varepsilon_{x_1,x_2} &= \PP(X=x_1)^\lambda \PP(X=x_2)^{1-\lambda} \\ \nonumber
 &\times \int [\PP(Y|X=x_1)]^\lambda [\PP(Y|X=x_2)]^{1-\lambda} dY,
\end{align} 
When the distributions are Gaussian, i.e.,  $\PP(Y|X=x_1) \sim \mathcal{N}(\mu_1,\sigma^2_1) $ and $\PP(Y|X=x_2) \sim \mathcal{N}(\mu_2,\sigma^2_2)$, the integration in \eqref{eq:chernoff-bound} can be obtained analytically as follows 
\begin{equation}
 \int [\PP(Y|X=x_1)]^\lambda [\PP(Y|X=x_2)]^{1-\lambda} dY = e^{-g(\lambda)} = \varepsilon'_{x_1,x_2},
\end{equation} 
where $g(\lambda)$ is called the Chernoff distance \cite{fukunaga2013introduction}:
\begin{align}
\label{eq:chernoff-distance}
\nonumber
    g(\lambda) = &\frac{\lambda(1-\lambda)}{2}(\mu_2-\mu_1)^2[\lambda \sigma_1^2 + (1-\lambda)\sigma_2^2]^{-1}   \\ 
    &+ \frac{1}{2} \ln\bigg[ \frac{|\lambda \sigma_1^2 + (1-\lambda)\sigma_2^2|}{|\sigma_1^2|^\lambda |\sigma_2^2|^{1-\lambda} }  \bigg].
\end{align}
The optimal transmit signals that minimizes the pairwise error probability can then be obtained by searching for the values of $\lambda$ and $(x_1, x_2)$ that maximizes the Chernoff distance in \eqref{eq:chernoff-distance} or minimizes $\varepsilon'_{x_1,x_2} = \exp[-g(\lambda)]$. 

For quadrature modulation with four transmit signals, i.e., $x_1 < x_2 < x_3 < x_4$, this optimization is performed by using the following compound design metric, 
\begin{equation}
\label{eq:minimize_chernoff}
    \varepsilon'_{x_1,x_2} + \varepsilon'_{x_2,x_3} + \varepsilon'_{x_3,x_4}.  
\end{equation} 
Note that the input $X$ and the observation $Y$ correspond to $\alpha$ and $\hat{\alpha}$ in Q-RSK, respectively, and to $c$ and $n_B$ in Q-CSK, respectively. 

\subsection{Decision Thresholds}
\label{sec:detection}
Let $H_m$ be the hypothesis that the symbol $m \in \{0,1,2,3\}$ is transmitted at the beginning of the $k^\text{th}$ signaling interval, and $Z_k$ be the corresponding received signal. As the received signals for both RSK and CSK are approximated as Gaussian distributed (see Section \ref{sec:stat_param}), the probability of observing $Z_k$ given that the hypothesis $H_m$ is true for the $k^\text{th}$ signaling interval can be given as 
\begin{equation}
    P(Z_k | H_m) = \frac{1}{\sqrt{2 \pi \sigma^2_m}} e^{\frac{(Z_k-\mu_m)^2}{2 \sigma^2_m}},
\end{equation} 
where $\mu_m$ and $\sigma^2_m$ are the mean and variance of the received signal corresponding to $m^\text{th}$ symbol of Q-RSK or Q-CSK. For a receiver employing maximum likelihood (ML) detection, the decision rule can be expressed as follows 
\begin{equation}
    \hat{m}_k = \underset{m}{\operatorname{arg~max}}~\PP(Z_k | H_m),    
\end{equation}
where $\hat{m}_k$ refers to the decided symbol for the $k^\text{th}$ signaling interval. ML approach utilized in the detection mechanism divides the entire range of received signal into $M = 4$ decision regions each of which correspond to a different symbol. The decision regions for the transmitted symbol $m$ can be defined as given below
\begin{align}
    D_m &= \{Z_k : \PP(Z_k |H_m) > \PP(Z_k |H_j)
     \forall j \neq m\}.
\end{align} 
The signal value where the conditional probability distributions of two adjacent symbols, $m-1$ and $m$, intersect will be the optimal decision threshold between them, $\lambda_m$, i.e., 
\begin{align}
\label{eq:threshold1}
\nonumber 
      \frac{1}{\sqrt{2 \pi \sigma^2_m}} e^{\frac{(\lambda_m-\mu_m)^2}{2 \sigma^2_m}} &= \frac{1}{\sqrt{2 \pi \sigma^2_{m-1}}} e^{\frac{(\lambda_m-\mu_{m-1})^2}{2 \sigma^2_{m-1}}} \\ 
      &\hspace{1.2cm}\text{for} \hspace{1mm} m = 1,2,3.
\end{align}
Solving \eqref{eq:threshold1}, we obtain the optimal decision thresholds: 
\begin{align}
\label{eq:threshold2}
\nonumber
 \lambda_m = &\frac{1}{\sigma^2_{m} -\sigma^2_{{m-1}}}
 (\sigma^2_{m} \mu_{{m-1}} - \sigma^2_{m-1} \mu_{{m}} +\sigma_{m} \sigma_{m-1} \\ \nonumber 
 &\times \sqrt{(\mu_{m} - \mu_{{m-1}})^2 + 2( \sigma^2_{m} - \sigma^2_{m-1} )\ln\Bigl(\frac{\sigma_{m}}{\sigma_{{m-1}}}}\Bigr) \\ 
 &\hspace{4cm} \text{for} \hspace{1mm} m = 1,2,3.
\end{align}

\subsection{Symbol Error Probability}
\label{sep}
Symbol error probability (SEP) is the probability of $m_k \neq \hat{m_k}$ where $m_k$ is the transmitted symbol, and $\hat{m}_k$ is the symbol decoded by the receiver in the $k^\text{th}$ signaling interval. The probability of erroneous detection can be computed as follows
\begin{equation}
    P(e|H_m) =  \underset{z \notin D_m}\int P(z|H_m)dz,
\end{equation} 
and by assuming that all the symbols have equal probability of being transmitted, SEP can be calculated by taking its mean over the entire symbol set as follows 
\begin{align}
\label{eq:RSK_Pe}
\nonumber
P_e   &=\frac{1}{4} \sum_{m=0}^{3} P(e | H_m) \\ \nonumber
&=\frac{1}{8} \bigg[ \mathrm{erfc}\bigg(\frac{\lambda_1-\mu_{0}}{\sigma_{0}\sqrt{2}}\bigg) +\mathrm{erfc}\bigg(\frac{\mu_{{3}}-\lambda_{3}}{\sigma_{{3}}\sqrt{2}}\bigg)  \\ 
\nonumber
&+\sum_{m=1}^{2}\bigg(\mathrm{erfc}\bigg(\frac{\mu_{{m}}-\lambda_{m}}{\sigma_{{m}}\sqrt{2}}\bigg) +\mathrm{erfc}\bigg(\frac{\lambda_{m+1}-\mu_{m}}{\sigma_{m}\sqrt{2}}\bigg)\bigg)\bigg]  \\ 
&\hspace{4.5 cm} \text{for} \hspace{1 mm} m=0,1,2,3,
\end{align} 
where $\mathrm{erfc}(z)= \frac{2}{\sqrt{\pi}} \int_{z}^{\infty} e^{-y^2}dy$ is the complementary error function.
\subsection{Numerical Results}
Here, we present numerical results on the analysis of the SEP performance of RSK and CSK in a mobile MC setting where both the transmitter and receiver are mobile. We provide the results obtained through analytical derivations and Monte Carlo simulations for various system settings. To investigate the impact of each parameter on performance, we conduct simulations where the transmitter sends $1000$ consecutive messages with a pre-defined signaling interval length $T_s$, while both the transmitter and receiver move in the channel following a random walk model. We assume that both the transmitter and the receiver know only the initial transmitter-receiver distance before they start the random walk.

The SEP for a single run of the simulation is obtained by calculating the average error probability observed during the transmission of the $1000$ messages. Simulations for each parameter are repeated $10^5$ times, and the average SEP over $10^5$ runs of the simulation is provided in the figures. The default values of the system parameters used in the analyses are given in Table \ref{tab:my-table}.

\begin{table}
\caption{Default Values of Simulation Parameters
}
\label{tab:my-table}
\begin{tabular}{|l|l|}
\hline
\multicolumn{1}{|c|}{\textbf{Parameter}}           & \multicolumn{1}{c|}{\textbf{Value}} \\ \hline
\multicolumn{1}{|c|}{Number of receptors ($N_R$)} & \multicolumn{1}{c|}{1000}  
 \\ \hline
\multicolumn{1}{|c|}{Diffusion coefficient of ligands ($D$)} & \multicolumn{1}{c|}{100 $\mu$ m$^2$/s }
\\ \hline
\multicolumn{1}{|c|}{Diffusion coefficient of Tx and Rx ($D_{tx,rx}$)} & \multicolumn{1}{c|}{0.001 $\mu$ m$^2$/s}
\\ \hline
\multicolumn{1}{|c|}{Initial Tx-Rx distance } & \multicolumn{1}{c|}{25 $\mu$ m}
\\ \hline
\multicolumn{1}{|c|}{Binding rate of ligands ($k^+$)} & \multicolumn{1}{c|}{20 $\mu$ m$^3$/s}  
\\ \hline
\multicolumn{1}{|c|}{Unbinding rate of type-1 ligands ($k_1^-$)} & \multicolumn{1}{c|}{10 s$^{-1}$}  
\\ \hline
\multicolumn{1}{|c|}{Unbinding rate of type-2 ligands ($k_2^-$)} & \multicolumn{1}{c|}{5 s$^{-1}$}  
\\ \hline
\multicolumn{1}{|c|}{Ligand similarity parameter ($\gamma$) } & \multicolumn{1}{c|}{2}
\\ \hline
\multicolumn{1}{|c|}{Signaling interval length ($T_s$) } & \multicolumn{1}{c|}{60 s} 
\\ \hline
\multicolumn{1}{|c|}{Disassociation constant of type-1 ligands ($K_{D,1}$)} & \multicolumn{1}{c|}{0.5 $\mu$ m$^{-3}$}  
\\ \hline
\multicolumn{1}{|c|}{Disassociation constant of type-2 ligands ($K_{D,2}$)} & \multicolumn{1}{c|}{0.25 $\mu$ m$^{-3}$}  
\\ \hline
\multicolumn{1}{|c|}{Maximum transmit power } & \multicolumn{1}{c|}{$5 \times K_{D,1}$} 
 \\ \hline
\end{tabular}
\end{table}

\begin{figure*}[t]
	\centering
	\begin{subfigure}[b]{0.32\textwidth}
		\centering
		\includegraphics[width=\textwidth]{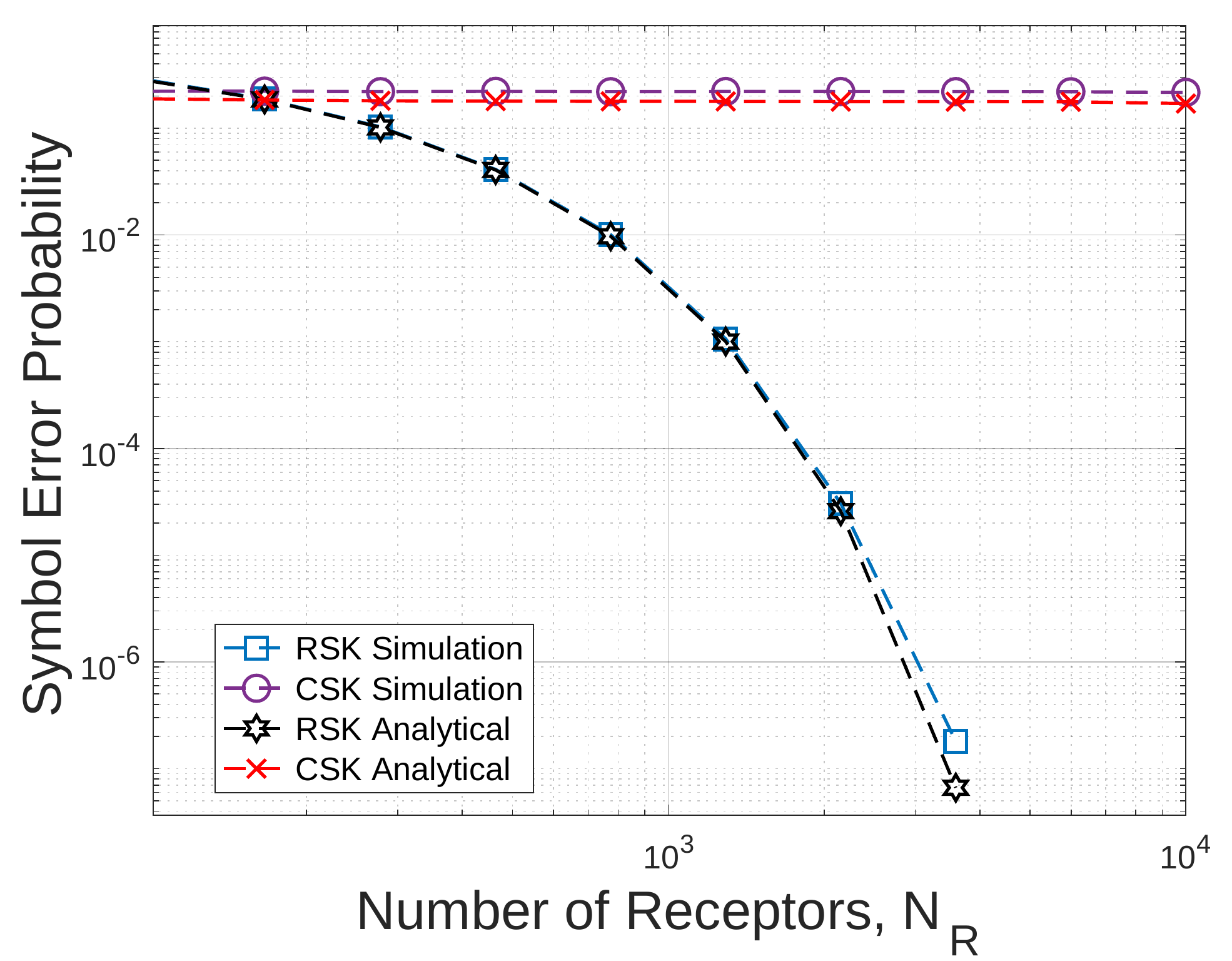}
		\caption{}
		\label{fig:mobil_receptor_number}
	\end{subfigure}
	\hfill
	\begin{subfigure}[b]{0.32\textwidth}
		\centering
		\includegraphics[width=\textwidth]{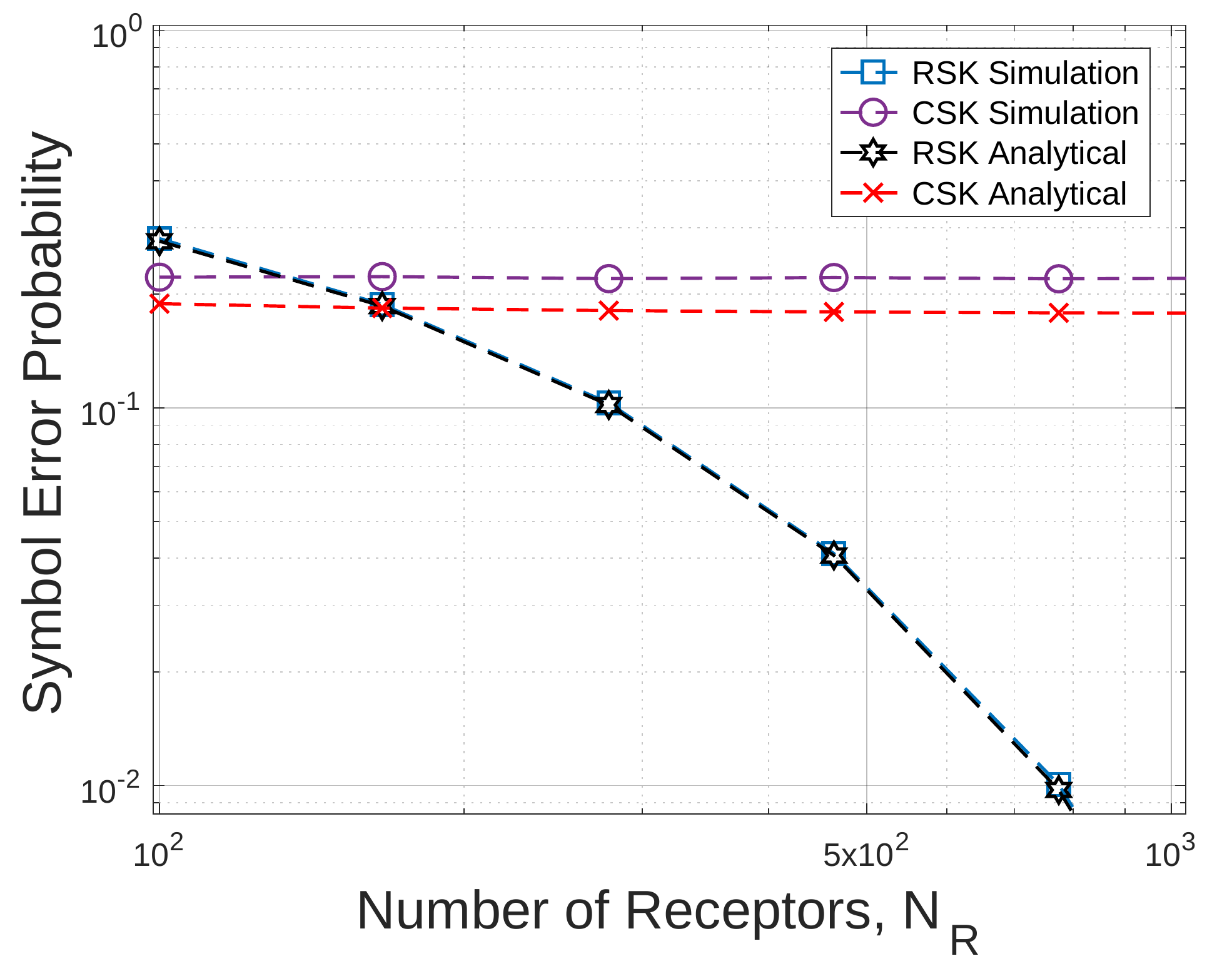}
		\caption{}
		\label{fig:mobil_receptor_number_close_3}
	\end{subfigure}
	\hfill
	\begin{subfigure}[b]{0.32\textwidth}
		\centering
		\includegraphics[width=\textwidth]{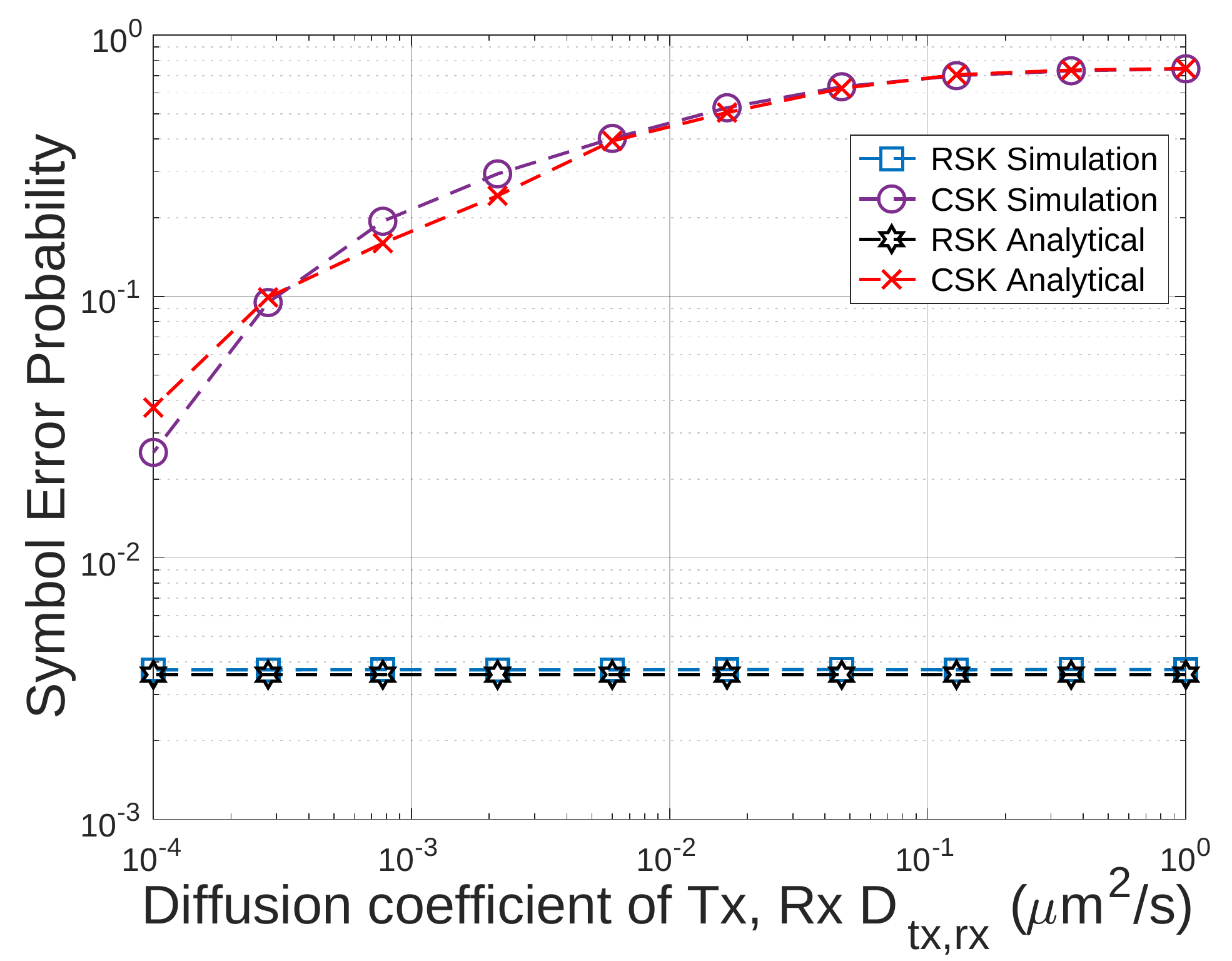}
		\caption{}
		\label{fig:mobil_diff}
	\end{subfigure}
	\caption{Symbol error probability of (a) RSK and CSK for both analytical and simulation results as a function of number of receptors, $N_R$, smaller region is redrawn on (b) for better visualization, (c) and for simulation results as a function of the diffusion coefficient of Tx and Rx, $D_{tx,rx}$.}
\end{figure*}

\subsubsection{\textbf{Effect of Number of Receptors}}
The number of receptors determines the number of independent samples taken for the estimation of the ligand concentration for CSK, and the ligand concentration ratio for RSK. As explained in Section \ref{sec:stat-rec-lig-conc}, statistics of the received ligand concentrations are predominantly determined by the mobility characteristics of the Tx-Rx pair. Hence, the effect of increasing number of independent samples is dwarfed by the impact of mobility in the case of CSK. Consequently, as observed in Figs. \ref{fig:mobil_receptor_number} and \ref{fig:mobil_receptor_number_close_3}, SEP for CSK is almost independent of the number of receptors in mobile MC, unlike the analysis on channel capacity for static MC. However, since the ligand concentration ratio is unaffected by the mobility of Tx and Rx, SEP for RSK decreases with increasing number of receptors, similar to the analysis on channel capacity in static MC case. Although CSK provides slightly better results than RSK when $N_R < 100$, RSK outperforms CSK significantly with a sufficient number of independent samples. Moreover, it can also be seen that the analytical results approximate the simulation results very accurately.

\begin{figure*}[t]
	\centering
	\begin{subfigure}[b]{0.41\textwidth}
		\centering
		\includegraphics[width=\textwidth]{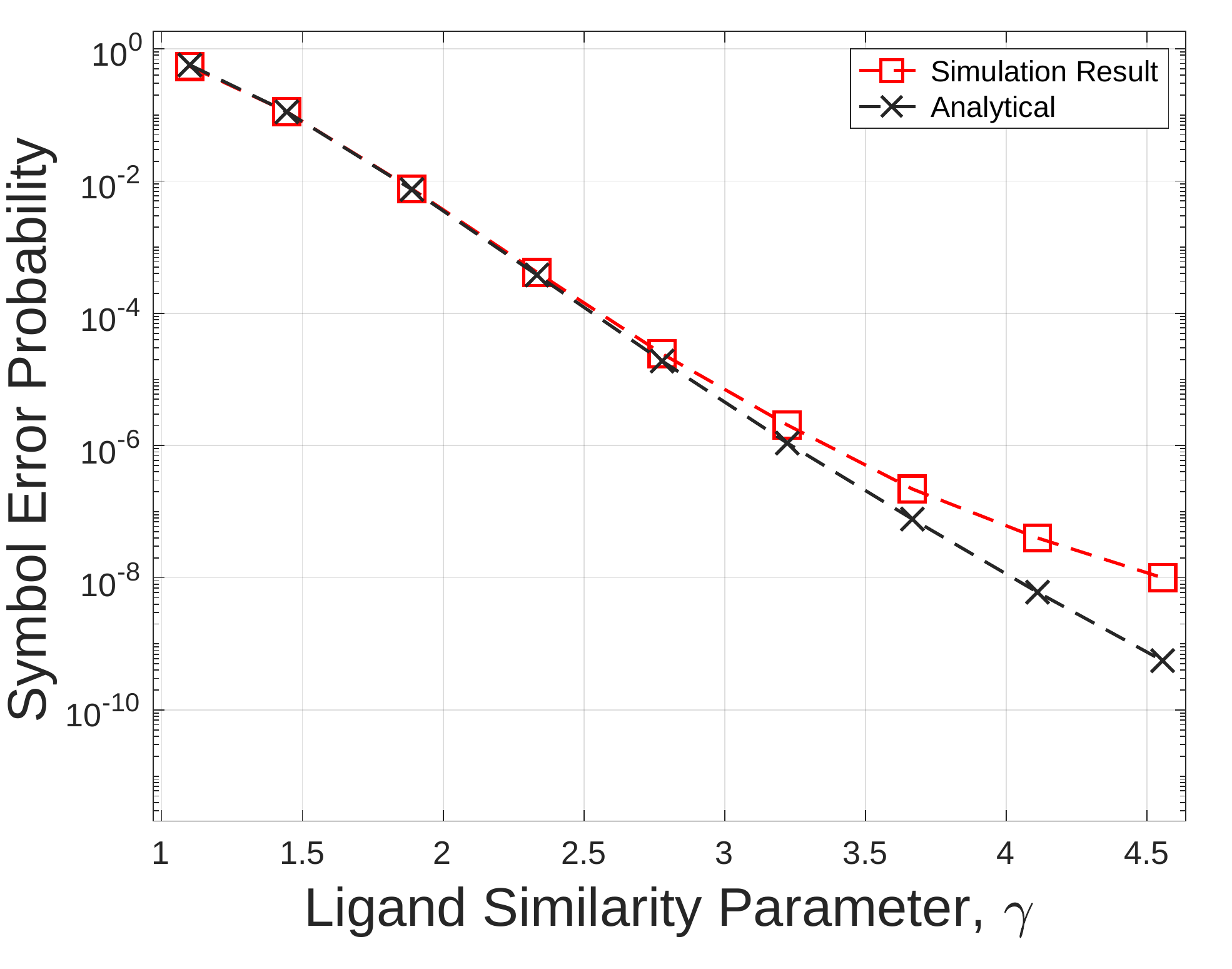}
		\caption{}
		\label{fig:similarity}
	\end{subfigure}
	\quad
	\begin{subfigure}[b]{0.41\textwidth}
		\centering
		\includegraphics[width=\textwidth]{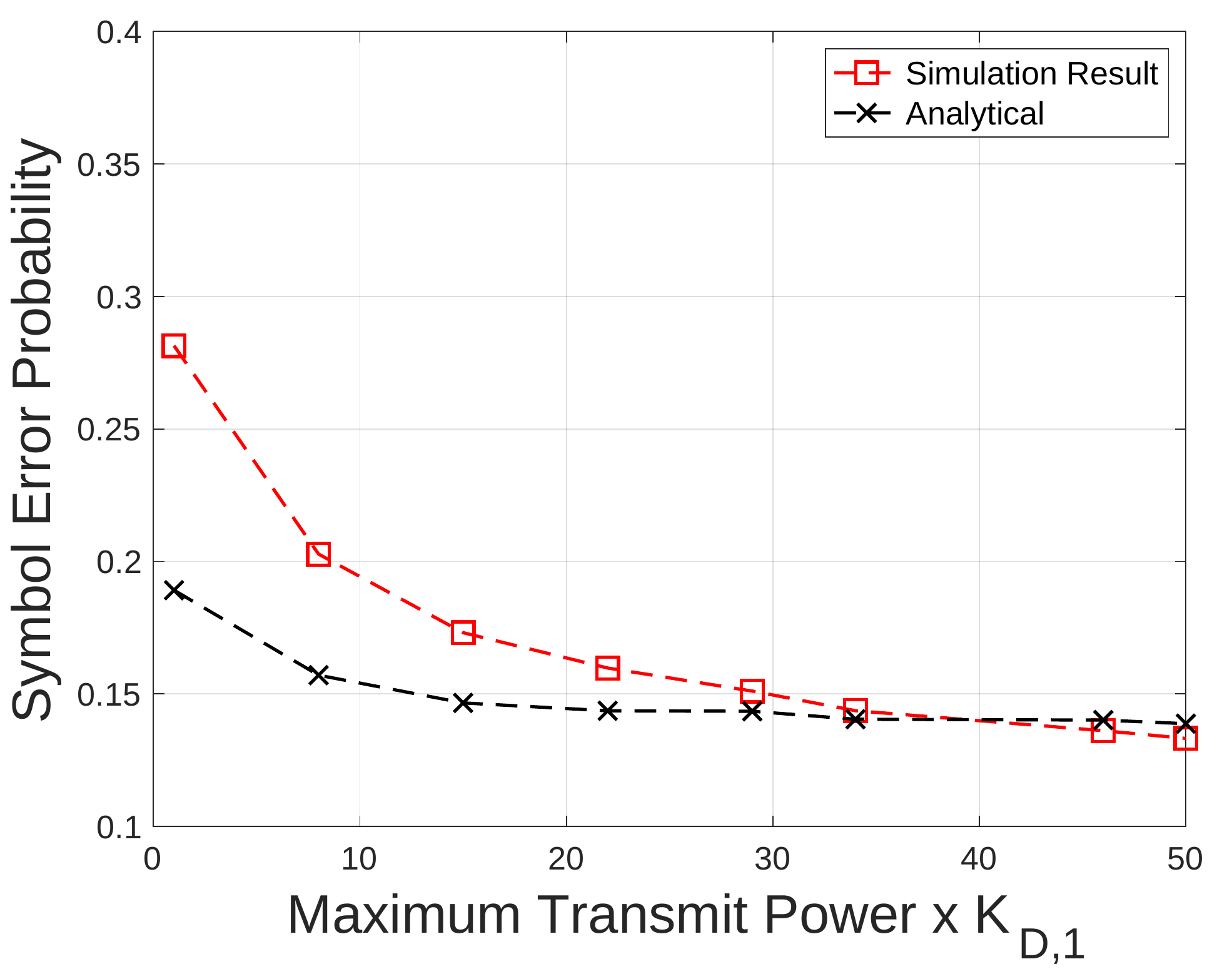}
		\caption{}
		\label{fig:mobilpowerlimited}
	\end{subfigure}
	\caption{Symbol error probability (a) for RSK as a function of similarity parameter, $\gamma$, and (b) for CSK as a function of maximum transmit power.}
\end{figure*}

\subsubsection{\textbf{Effect of Diffusion Coefficient of Transmitter and Receiver}}
We investigate the impact of transmitter and receiver mobility on RSK and CSK performance by tuning their diffusion coefficient, i.e., $D_{tx,rx}$. In this analysis, we keep the diffusion coefficient of the ligands constant at $D = 100 \mu m^2/s$, whereas $D_{tx,rx}$ is varied from $D \times 10^{-6} = 10^{-4} \mu m^2/s$ to $D\times 10^{-2} = 1 \mu m^2/s$. As shown in Fig. \ref{fig:mobil_diff}, at low mobility conditions with small $D_{tx,rx}$, CSK outperforms RSK. However, with increasing mobility, the variance of the Tx-Rx distance at any time increases, leading to higher fluctuations on the ligand concentration in the vicinity of the receiver, resulting in a drastic increase of SEP for CSK. Since the proportional changes in the concentrations of individual ligands are the same, the concentration ratio of the two ligand types used in RSK is unaffected by the variations of $D_{tx,rx}$. Consequently, mobility characteristics do not affect the concentration ratio, ensuring that the error performance of RSK remains unaffected. In contrast,  the CSK performance is highly correlated with mobility, reaffirming the advantage of RSK in time-varying MC channels.

\subsubsection{\textbf{Effect of Similarity between Ligand Types on RSK Performance}}
The similarity between type-1 and type-2 ligands used for RSK in terms of their affinity with the receptors has substantial impact on SEP, as demonstrated in Fig. \ref{fig:similarity}. Increasing $\gamma$, which reflects a decrease in similarity, improves the ability of the suboptimal estimator to distinguish between the two types of ligands based on the bound time duration of receptors. Fig. \ref{fig:similarity} shows that SEP is close to $0.75$, where ligands are very similar to each other, i.e., $\gamma \approx 1$. However, as $\gamma$ increases, SEP decreases drastically, and for $\gamma \approx 4.5$, it attains values on the order of $10^{-8}$, which practically corresponds to a no-error case in a mobile MC scenario. Therefore, RSK can provide significantly low SEP with the careful choice of ligand types.

\subsubsection{\textbf{Effect of Maximum Transmit Power on CSK Performance}}
For CSK, we analyze the effect of the maximum transmit power which may need to be set given a transmitter with limited molecule reservoir. Here, for convenience of the analysis, the maximum transmit power is given in terms of the maximum received concentration, which corresponds to the maximum number of transmitted molecules scaled by the CIR of the channel at the initial transmitter and receiver position. Note that a similar analysis has been done in Section \ref{sec:capacity_result} by investigating the effect of maximum received concentration over the channel capacity. Similar to the capacity analysis, SEP for CSK decreases with the increasing maximum transmit power in a mobile MC channel, as is clear in Fig. \ref{fig:mobilpowerlimited}. This result is expected for CSK since expanding the input concentration range (input space) increases the distance between the transmit signals in terms of ligand concentration, improving the distinguishability of transmitted symbols at the receiver, thereby lowering SEP.

\section{Conclusion}

We performed an information-theoretical analysis of the MC channel with RSK modulation considering two different ratio estimation schemes, varying in their optimality and complexity, for a ligand-receptor-based receiver. Additionally, we analyzed a practical time-varying MC case where both the receiver and transmitter are mobile. Performance of RSK has been numerically compared to that of CSK modulation in terms of the corresponding channel capacity and symbol error probability. The results demonstrated that RSK modulation outperforms CSK modulation particularly when the transmitter and receiver are mobile, or the transmitter is power-limited, such that the received ligand concentration is time-varying or upper-bounded. Future work will focus on the analysis of RSK modulation in scenarios involving multiple transmitters and receivers in time-varying channel conditions.

\appendices
\section{On the Intersymbol Interference}
\label{AppendixA}

For the default simulation parameters provided in Table \ref{tab:my-table}, the expected value and the variance of peak time distribution becomes 3.7\% and 4.37\% of the signaling interval, respectively. Since the received concentration decreases rapidly after $\tau_{peak}$ in a signaling interval, and the sum of the mean and variance of the peak time distribution is less than \%10 of the signaling interval, we can safely assume that previous transmission will not affect the current transmission. Hence we neglected ISI in our analyses. Nevertheless, we provide an analysis for the effect of the sampling interval $T_S$ to verify that ISI can be neglected.

\begin{figure}[ht!]
  \includegraphics[width=\linewidth]{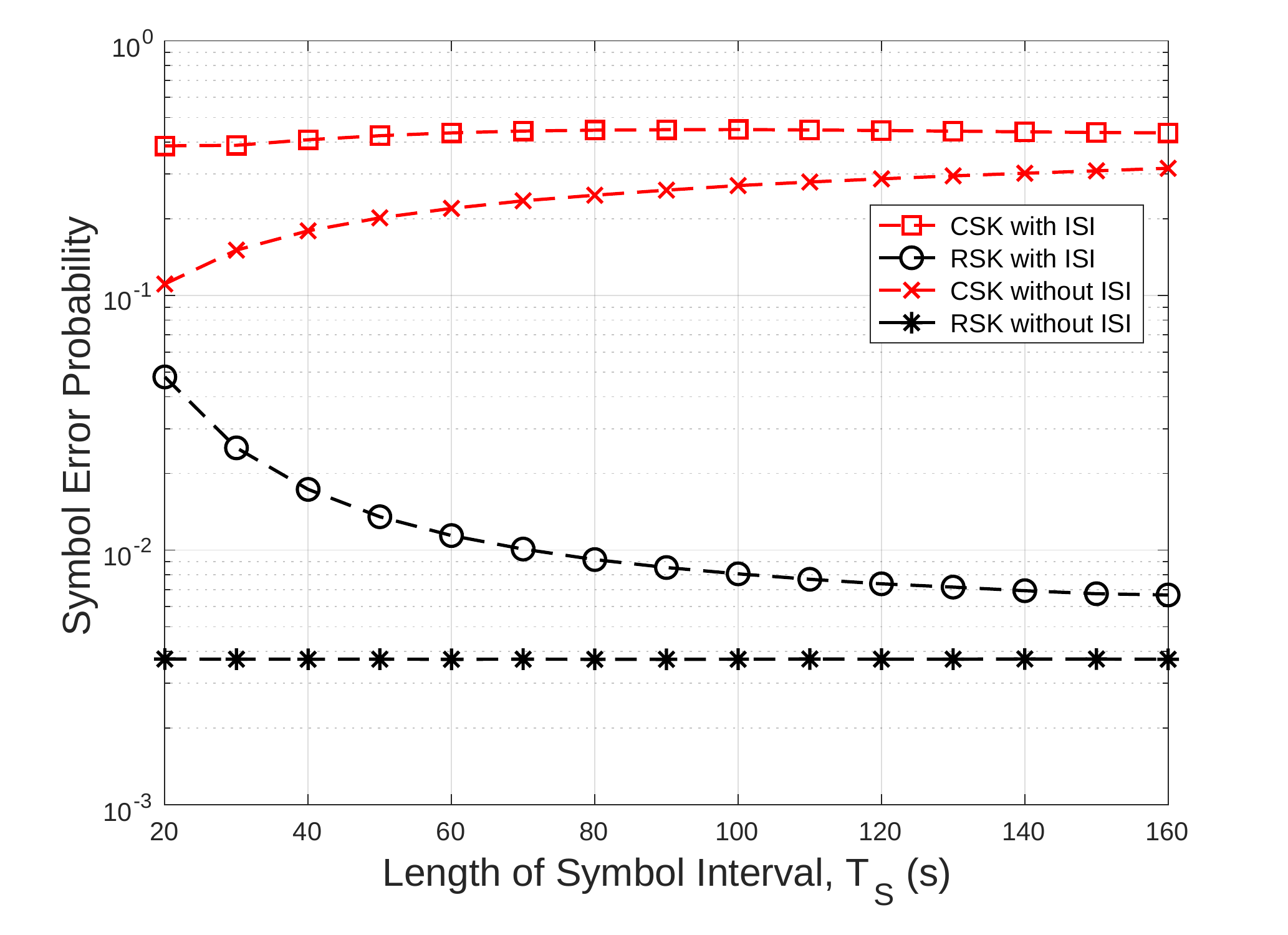}
  \caption{Symbol error probability of RSK and CSK as a function of varying signaling interval length for both cases with and without ISI.}
  \label{fig:isi_noisi}
\end{figure}

As is seen in Fig. \ref{fig:isi_noisi} for both RSK and CSK modulations, differences in the SEP between the cases with and without ISI are in an acceptable tolerance band for $T_S > 50$s. Therefore, we conclude that ISI can be neglected for $T_S > 50$s. By setting $T_S = 60$s, we conducted our analyses in this paper without considering ISI.

\bibliographystyle{IEEEtran}

\bibliography{references}

\newpage

\vfill

\end{document}